\documentclass{article}
\usepackage{graphicx}
\usepackage{amsmath,amssymb}
\usepackage{graphics,color,array,calc,rotating,epsfig,psfrag}
\numberwithin{equation}{section}
\usepackage{cite}
\usepackage{bm}
\usepackage{dcolumn}  
\usepackage{amsfonts}
\usepackage[mathscr]{eucal}
\def\be{\begin{equation}} \def\ee{\end{equation}}
\def\bea{\begin{eqnarray}} \def\eea{\end{eqnarray}}
\def\bs{\boldsymbol}

\newcommand\prt{\partial}

\newcommand{\nn}{\nonumber}
\begin{document}
\baselineskip 18pt%
\begin{titlepage}
\vspace*{1mm}%
\hfill%
\vspace*{15mm}%
\hfill
\vbox{
    \halign{#\hfil         \cr
          } 
      }  
\vspace*{20mm}

\centerline{{\Large {\bf On asymptotically $AdS$-like solutions of three dimensional massive gravity}}}
\vspace*{5mm}
\begin{center}
{ Ahmad Ghodsi\footnote{ahmad@ipm.ir, a-ghodsi@ferdowsi.um.ac.ir} and Davood Mahdavian Yekta \footnote{da.\,mahdavianyekta@stu-mail.um.ac.ir}}\\
\vspace*{0.2cm}
{ Department of Physics, Ferdowsi University of Mashhad, \\
P.O. Box 1436, Mashhad, Iran}\\
\vspace*{0.1cm}
\end{center}

\begin{abstract} 
In this paper we have added Maxwell, Maxwell-Chern-Simons and gravitational Chern-Simons terms to  Born-Infeld extended new massive gravity and we have found different types of (non)extremal charged black holes. For each black hole we find mass, angular momentum, entropy and temperature. Since our solutions are asymptotically $AdS$ or warped-$AdS$, we infer central charges of dual CFTs by using Cardy's formula. Computing conserved charges associated to asymptotic symmetry transformations confirms  calculation of central charges. 
For CFTs dual to asymptotically $AdS$ solutions we find left central charges from Cardy's formula, while conserved charge approach gives  both left and right central charges. For  CFTs dual to asymptotically warped-$AdS$ solutions, left and right central charges are equal when we have Maxwell-Chern-Simons term but they have different values when gravitational Chern-Simons term is included. 
\end{abstract} 

\end{titlepage}

\section{Introduction}
Three dimensional massive gravity is an interesting subject in studying gravity. Topological Massive Gravity (TMG) is obtained by adding a gravitational Chern-Simons term to pure gravity \cite{Deser:1982vy}. New Massive gravity (NMG) on the other hand is constructed by adding higher order curvature terms to the Einstein-Hilbert term \cite{Bergshoeff:2009hq}. A combination of both theories, known as General Massive Gravity (GMG), is also  studied in \cite{Bergshoeff:2009tb}.

The mentioned theories have different properties. For example TMG theory, due to the presence of a Chern-Simons term, is a parity-violating gravitational model  while  NMG theory is constructed out of parity-preserving terms. 

Several solutions such as the BTZ black holes, Warped-$AdS$ background, logarithmic and polynomial solutions have been found for these gravitational theories \cite{Banados:1992wn}-\cite{Ahmedov:2010uk}. To obtain charged solutions for these theories, one may add gauge fields through the Maxwell term. This leads to TMGE, which has been studied in \cite{Moussa:2008sj} and  NMGE \cite{Ghodsi:2010gk}.

Since massive gravity theories in three dimensions have different sectors, for example they contain solutions which are asymptotically $AdS_3$ or  asymptotically warped-$AdS_3$, it will be very interesting to study the $AdS_3/CFT_2$ correspondence in this context (see \cite{Aharony:1999ti}). For example several attempts have been done by studying isometery groups and computing the central charges of asymptotic algebras \cite{BH}-\cite{Maldacena:1998bw}. In this paper we try to find different solutions for different combinations of massive gravities. We will choose two directions to extend  massive gravity theories. In one direction we add gauge fields and in other direction we consider higher order curvature terms. These extensions enable us to get more possible solutions to study the $AdS_3/CFT_2$ correspondence. We use the following approach for these extensions of massive gravities. 

NMG contains curvature square terms. To extend this theory in order to have higher order curvature terms, several attempts have been done. One of the recent developments is Born-Infeld extension of new massive gravity \cite{Gullu:2010pc}
\bea\label{BINMG0}
S=\frac{2m^2}{\kappa}\int d^3 x\Big[\sqrt{-det(g_{\mu\nu}+\frac{\sigma}{m^2}\,G_{\mu\nu})}-(1+\frac{\Lambda}{2m^2})\sqrt{-det g_{\mu\nu}}\Big]\,,
\eea
where $g_{\mu\nu}$ is metric and $G_{\mu\nu}={\cal R}_{\mu\nu}-\frac12 {\cal R} g_{\mu\nu}$ is Einstein tensor.  The parameter $m$ is a mass parameter and $\kappa=8\pi G_3$ is the three dimensional gravitational constant. To have a positive coefficient for scalar curvature we choose $\sigma=-1$. This condition guaranties  unitarity of the theory \cite{Gullu:2009vy}.

Gauge field strength and  Chern-Simons term have been added in \cite{Ghodsi:2010ev} in order to study charged solutions 
\bea\label{BINMG}
S=\frac{2m^2}{\kappa}\int d^3 x\Big[\sqrt{-det(g_{\mu\nu}+\frac{\sigma}{m^2}\,G_{\mu\nu}+ a F_{\mu\nu})}-(1+\frac{\Lambda}{2m^2})\sqrt{-det g_{\mu\nu}}\,\Big]
+\frac{\mu}{2}\int d^3 x\,\epsilon^{\mu\nu\rho} A_{\mu}\partial_{\nu}A_{\rho}\,,
\eea
where $a$ and $\mu$ are two constants and $F_{\mu\nu}=\prt_{\mu} A_{\nu}-\prt_{\nu} A_{\mu}$ is the field strength of a $U(1)$ gauge field. In \cite{Ghodsi:2010ev} we have studied warped-AdS solution for this theory. In this paper we try to find other possible solutions. We also extend this action and add topological gravitational Chern-Simons term.

If we insert $a\!=\!\mu\!=\!0$ and only consider expansion up to second order of curvature, one will find the new massive gravity action \cite{Bergshoeff:2009hq}. Expansion to  next leading order terms, gives deformation of NMG obtained by AdS/CFT correspondence consideration \cite{Sinha:2010ai}. Uncharged $AdS$ black hole solutions for this  action have been found in \cite{Alishahiha:2010iq}. 

This paper is organized as follows: In section two we will see how one can find solutions for equations of motion. First we expand Lagrangian (\ref{BINMG}) up to cubic curvature terms and then we introduce the dimensional reduction procedure. We discuss mass and angular momentum of solutions. We show how to find some useful thermodynamical quantities such as entropy and Hawking temperature. In section three we obtain new charged solutions and compute their mass, angular momentum and entropy. In section four we add gravitational Chern-Simons term to the Lagrangian and find different extremal and non-extremal solutions. In section five we calculate central charges of two dimensional conformal field theories dual to the black hole solutions, by using the Cardy's formula. Sections six and seven contain attempts to find central charges by computing the conserved charges associated to  asymptotic symmetry transformations. In last section we summarize our results.
\section{How to find solutions and their physical quantities?}
\subsection{The Lagrangian}
In this paper we are interested in physical properties of the solutions corresponding to the expansion of Lagrangian (\ref{BINMG}). Expansion of (\ref{BINMG}) up to second, forth and sixth orders of derivatives gives the following  Lagrangians at different orders 
\bea\label{EXL}
\mathcal{L}_{O(2)}
&=&\frac{2m^2}{\kappa}\,\sqrt{-g}\Big[\frac{1}{4m^2}\,R-\frac14\,a^2\,Tr(F^2)\Big]\,-\sqrt{-g}\,\frac{\Lambda}{\kappa}\,,\nn\\
\mathcal{L}_{O(4)}
&=&\frac{2m^2}{\kappa}\,\sqrt{-g}\Big[-\frac{1}{4m^4}\Big(Tr(R^2)-\frac38\,R^2\Big)-\frac{ a^2}{2m^2}\Big(Tr(RF^2)-\frac38\,R\,Tr(F^2)\Big)\nn\\
&-&\frac18\, a^4\Big(Tr(F^4)-\frac14(Tr(F^2))^2\Big)\Big]\,,\nn\\
\mathcal{L}_{O(6)}
&=&\frac{2m^2}{\kappa}\,\sqrt{-g}\Big[-\frac{1}{6m^6}\Big(Tr(R^3)-\frac98\,R\,Tr(R^2)+\frac{17}{64}R^3\Big)\nn\\
&-&\frac{a^2}{m^4}\Big(\frac34\,Tr(R^2F^2)-\frac58\,R\,Tr(RF^2)
+\frac{19}{128}\,R^2\,Tr(F^2)-\frac{1}{16}Tr(R^2)Tr(F^2)\Big)\nn\\
&+&\frac{ a^4}{2m^2}\Big(Tr(RF^4)-\frac{7}{16}RTr(F^4)+\frac{7}{64}R(Tr(F^2))^2-\frac14 Tr(RF^2)Tr(F^2)\Big)\nn\\
&-&\frac{a^6}{12}\Big(Tr(F^6)-\frac38 Tr(F^2)Tr(F^4)+\frac{1}{32}(Tr(F^2))^3\Big)\Big]\,,
\eea
where $Tr(AB)=A_{\mu\nu}B^{\nu\mu}$. We also add Maxwell-Chern-Simons term
\be\label{csl}
\mathcal{L}_{CS}
=\frac{\mu}{2}\,\epsilon^{\mu\nu\rho}A_{\mu}\partial_{\nu}A_{\rho}\,.
\ee
It is obvious that if we insert $a=\mu=0$ then we will find extended NMG Lagrangian \cite{Gullu:2010pc},\cite{Sinha:2010ai}. 
We also consider gravitational Chern-Simons Lagrangian 
\be\label{LTMG}
\mathcal{L}_{GCS}=\frac{1}{4\kappa\,\mu_{G}}\, \sqrt{-g}\,\epsilon^{\lambda\mu\nu}\,\Gamma^{\delta}_{\lambda\sigma}\,\Big(\prt_{\mu}\Gamma^{\sigma}_{\delta\nu}+\frac23\,\Gamma^{\sigma}_{\mu\tau}\,\Gamma^{\tau}_{\nu\delta}\Big)\,.
\ee
To write Maxwell Lagrangian in its canonical form from now on we consider $a^2=-\frac{\kappa}{2m^2}$.

\subsection{Ansatz}
In this paper we would like to find stationary rotationally symmetric solutions. To do this, the best way is to use the dimensional reduction procedure introduced by \cite{Clement:2009gq}, \cite{Moussa:2008sj}. In this procedure, one considers a three dimensional metric which has symmetry group of the $SL(2,R)$ transformations and gauge field has a $SL(2,R)$ doublet representation.  Therefore we can write the metric and gauge field as follows
\bea \label{ansatz}
ds^2=\lambda_{ab}(\rho)\,dx^a dx^b + \zeta^{-2}(\rho)R^{-2}(\rho) \,d\rho^2\,,
\qquad {\cal A} = A_a(\rho) \,dx^a\,,
\eea
where ($a,b=0,1$) and ($x^0 = t$, $x^1 = \varphi$). The components of $\lambda$ can be expressed by a $2 \times 2$ matrix
\be
\lambda = \left(
\begin{array}{cc}
T+X & Y \\
Y & T-X
\end{array}
\right)\,.
\ee
To obtain a solution we first insert the above ansatz into the Lagrangian and then we find equations of motion by the variation of Lagrangian with respect to $A_a, \zeta, T, X $ and $Y$.

To find the physical quantities of solutions such as temperature or angular velocity, it would be better to write the ansatz (\ref{ansatz}) in its $ADM$ form \cite{Arnowitt:1962hi} i.e.
\be 
ds^2\,=\,- N(r)^2\,dt^2+\,K(r)^2\,\big(d\phi + N^{\phi}\,dt\big)^2\,+\frac{r^2dr^2}{K(r)^2\,N(r)^2}\,, 
\ee
where we have used the following definitions
\be 
\label{para} N^2\,=\,\frac{R^2}{T - X}\,,\quad N^{\phi}\,=\,\frac{Y}{T - X}\,,\quad K^2\,=\,T - X\,,\quad r^2\,=\,2\,\rho\,,\quad \zeta(\rho)=1\,,
\ee
and $R^2=-T^2+X^2+Y^2$. 
\subsection{Mass and angular momentum}
To compute mass and angular momentum of a black hole we use the Clement's approach presented in \cite{Clement:2009gq}. In this approach there is a conserved current called super-angular momentum vector $\bs J$.  Angular momentum and  mass  are related to super-angular momentum via 
\be \label{ang} 
J=2\pi\,(\delta {\bs J}^T- \delta {\bs J}^X)\,,\quad M=2\pi\,(\delta{\bs J}^Y+\Delta)\,,
\ee 
where $\delta$ denotes the difference between values of super-angular momentum for solution and background. The background usually is a massless static solution. We will show that either $\Delta=0$ or $\Delta=\delta{\bs J}^Y$.

Since we have considered gauge fields, we need to extend the Clement's approach. This has been already done in \cite{Ghodsi:2010ev}.
The Lagrangian we are dealing with has a $SL(2,R)$ symmetry and super-angular momentum is its conserved current so under infinitesimal $SL(2,R)$ transformations we find the following field transformations 
\bea
&&\Delta T= \epsilon^1 Y-\epsilon^2 X\,,\quad
\Delta X= \epsilon^0 Y-\epsilon^2 T\,,\quad
\Delta Y=-\epsilon^0 X+\epsilon^1 T\,,\nonumber\\
&&\Delta A_0= \frac12(\epsilon^0+\epsilon^1) A_1-\frac12\epsilon^2 A_0\,,\quad
\Delta A_1= \frac12(-\epsilon^0+\epsilon^1) A_0+\frac12\epsilon^2 A_1\,,\quad \Delta A_2=0\,.
\eea
The conserved super-angular momentum current has two parts. For the gravity part we find
\bea \label{JGR}
\bs{J}_{Gr}=\Big[\!\!\!\!&+&\!\!\!\!(\frac{\partial L}{\partial X'}Y-\frac{\partial L}{\partial Y'}X)
-((\frac{\partial L}{\partial X''})'\,Y-(\frac{\partial L}{\partial Y''})'\,X)
+(\frac{\partial L}{\partial X''}Y'-\frac{\partial L}{\partial Y''}X'),\nonumber\\
&+&\!\!\!\!(\frac{\partial L}{\partial T'}Y\,+\frac{\partial L}{\partial Y'}T)
\,-((\frac{\partial L}{\partial T''})'\,Y\,+\,(\frac{\partial L}{\partial Y''})'\,T)
\,+(\frac{\partial L}{\partial T''}Y'\,+\frac{\partial L}{\partial Y''}T'),\nonumber\\
&-&\!\!\!\!(\frac{\partial L}{\partial T'}X+\frac{\partial L}{\partial X'}T)
+((\frac{\partial L}{\partial T''})'\,X\,+\,(\frac{\partial L}{\partial X''})'\,T)
-(\frac{\partial L}{\partial T''}X'+\frac{\partial L}{\partial X''}T')
\Big]\,,
\eea
where primes denote derivatives with respect to $\rho$. The above vector is equivalent to the vector current which has been found in \cite{Clement:2009gq}. For electromagnetic part one finds \cite{Ghodsi:2010ev}
\bea \label{JEM}
\bs{J}_{EM}=\frac12\Big[(\frac{\partial L}{\partial A_0'}A_1-\frac{\partial L}{\partial A_1'}A_0)\,,
(\frac{\partial L}{\partial A_0'}A_1+\frac{\partial L}{\partial A_1'}A_0)\,,
-(\frac{\partial L}{\partial A_0'}A_0-\frac{\partial L}{\partial A_1'}A_1)
\Big]\,.
\eea
The total super-angular momentum is the sum of these two parts, i.e. $\bs{J}=\bs{J}_{Gr}+\bs{J}_{EM}$.
\subsection{Thermodynamics}
In addition to mass, angular momentum or charge one can find thermodynamical properties of black holes. The most important parameter is entropy and its value is given by using the Wald's formula \cite{Wald:1993nt} 
\bea \label{ent1}
S_W &=& 4\pi A_h\Big(\frac{\delta{\cal L}}
{\delta {\cal R}_{0202}}(g^{00}g^{22})^{-1}\Big)_h\,,
\eea
where $A_h $ is the horizon's area. By adding the TMG Lagrangian we must also consider its contribution to entropy \cite{Moussa:2008sj}. So the total value of entropy in presence of the TMG Lagrangian will be
\be
\label{ent2} S=S_{W}-\frac{2\pi^2}{\kappa\,\mu_{G}}r_h^3\,(N^\varphi)'\,, 
\ee
where  $r_h$ is the location of horizon and it can be found from ADM form of the metric. 

To check the first law of thermodynamics for black holes we need to find more physical quantities.
According to definitions in (\ref{para}) one can read Hawking's temperature, angular velocity and area of horizon from ADM form of metric. These are
\be \label{temp} 
T_H=\frac{1}{4\pi\,}
\frac{\big[R^2\big]'}{\sqrt{T-X}}\Big{|}_{\rho_{\!h}}\,,\quad \Omega_{h}=-\frac{Y}{T-X}\Big{|}_{\rho_{\!h}}\,,\quad A_h=2\pi\sqrt{T-X}\,\Big{|}_{\rho_{\!h}}\,,
\ee
where prime is derivative with respect to $\rho$. Since we have a gauge field, we can also
find the value of electric potential $\Phi$ at  horizon 
\be \label{phih}
\Phi_h=-\big(A_t+\Omega_h\,A_\phi\big)\,.
\ee
We will check that black hole solutions satisfy the first law of thermodynamics i.e,
\be\label{firstlaw}
dM=T_h\, dS+\Omega_h\, dJ+ 2\pi \Phi_h dQ\,,
\ee 
where $Q$ is the electric charge for each charged solution.

There is another approach to find mass of
each black hole from the integrated form of the first law or the Smarr-like formula \cite{Smarr:1972kt}. Depending on each solution we find one of the following relations
\bea \label{MASS} 
M=T_H S\,+2\Omega_h J+\frac12\,\Phi_h \bar{Q}\,,\quad M=\frac12\,T_H S\,+\Omega_h J+\,\Phi_h \bar{Q}\,, \eea
where $\bar{Q}=2\pi Q$ . 
\section{Solutions of  extended NMG}
In this section we consider Lagrangian (\ref{EXL}) together with Chern-Simons term (\ref{csl}). We use ${\mathcal{O}}(2), {\mathcal{O}}(4)$ and ${\mathcal{O}}(6)$ notations for order of expansion of Born-Infeld Lagrangian and use ${\mathcal{O}}(\infty)$  for Born-Infeld Lagrangian. 
\subsection{MCS\,-\,charged solution}
To start, let's add Maxwell-Chern-Simons Lagrangian (\ref{csl}) to ${\mathcal{O}}(2), {\mathcal{O}}(4)$, ${\mathcal{O}}(6)$ and ${\mathcal{O}}(\infty)$ Lagrangians. We call the corresponding solutions, MCS-charged black holes.  We consider the following ansatz for  metric and  gauge field, which is a self-dual MCS solution found in \cite{Clement:1995zt}
\be 
\label{mcsx} {\bs X}={\bs \alpha}\,C \rho^{\nu}+{\bs \beta}\,\rho+{\bs \gamma} C_0\,,\quad{\cal A}=Q
\rho^\frac{\nu}{2}\,\big(dt-l
d\phi\big)\,, 
\ee 
where $Q$ is the electric (magnetic) charge of $U(1)$ gauge field. The constant parameters $C, C_0,\nu$ and $l$ can be computed from equations of motion.
In this ansatz $\bs \alpha$, $\bs \beta$ and $\bs \gamma$ are vectors which determine the frame of solutions. Equations of motion for $T,X$ and $Y$ restrict $\bs \alpha$ and $\bs \beta$ vectors to  ${\bs \alpha}^2={\bs \alpha}.{\bs \beta}=0$ and $\bs \gamma||\bs \alpha$. So we choose the following frame for self-dual solutions 
\be
{\bs \alpha}={\bs \gamma}=\Big(l^2+1\,,-l^2+1\,,-2l\Big)\,,\quad{\bs \beta}=\Big(\frac{l^2-1}{l^2}\,,-\frac{l^2+1}{l^2}\,,0\Big)\,. 
\ee
Inserting this frame into (\ref{mcsx}) and using Euler-Lagrange equations of motion for $T,X$ or $Y$ we can find $C$ as a function,  $C=C(Q,\nu,m,l)$. One can easily find $C$ as in the first column of table 2. In next step we insert the above ansatz into equations of motion for gauge field. These equations give two values for $\nu$. Either $\nu$ is a function as $\nu=-\mu l H(m,l)$ or $\nu=0$. 

Let us first start with nonzero value of $\nu$. Solving equations of motion give values of $H$ in the last column of table 2. The remaining equation of motion is for $\zeta$. This equation gives the value of  cosmological constant (see table 1). 
\begin{table}[ht]\renewcommand{\arraystretch}{1.5}
\center
\begin{tabular}{|c|c|}\hline
     & $\Lambda$  \\ \hline 
${\mathcal{O}}(2)$ & $-\frac{1}{l^2}$\\ \hline
${\mathcal{O}}(4)$ & $-\frac{1}{l^2}(1+\frac{1}{4 \xi^2})$\\ \hline
${\mathcal{O}}(6)$ & $-\frac{1}{l^2}\,(1+\frac{1}{4 \xi^2}+\frac{1}{8\xi^4})$\\ \hline
${\mathcal{O}}(\infty)$& $-\frac{2\xi^2}{l^2}(1-(1-\frac{1}{\xi^2})^\frac12) $\\ \hline
\end{tabular}
\caption{The value of cosmological constant $\Lambda$ in each order of Lagrangian ($\xi=m l$).}
\end{table}
\begin{table}[ht]\renewcommand{\arraystretch}{1.5}
\center
\begin{tabular}{|c|c|c|c|}\hline
     & $C$ & $C'$ & $H$  \\ \hline 
${\mathcal{O}}(2)$ & $-\frac{\kappa Q^2 \nu}{4(\nu-1)}$ & 1 & $1$\\ \hline
${\mathcal{O}}(4)$ & $-\frac{\kappa Q^2 \nu}{4(\nu-1)}\frac{2\xi^2+1+4\nu(\nu-1)}{2\xi^2-1-8\nu(\nu-1)}$ & $\big(1-\frac{1}{2  \xi^2}\big)$ & $\big(1+\frac{1}{2  \xi^2}\big)^{-1}$\\ \hline
${\mathcal{O}}(6)$ & $-\frac{\kappa Q^2 \nu}{4(\nu-1)}\frac{8\xi^4+4\xi^2+3+(16\xi^2+24)\nu(\nu-1)}{8\xi^4-4\xi^2-1-(32\xi^2+16)\nu(\nu-1)}$ & $\big(1-\frac{1}{2  \xi^2}-\frac{1}{8  \xi^4 }\big)$ & $(1+\frac{1}{2 \xi^2}+\frac{3}{8 \xi^4})^{-1}$\\ \hline
${\mathcal{O}}(\infty)$ & $-\frac{\kappa Q^2 \nu}{4(\nu-1)}\frac{\xi^2-1+2\nu(\nu-1)}{\xi^2-(2\nu-1)^2}\,(1-\frac{1}{\xi^2})^{-1}$ & $\big(1-\frac{1}{ \xi^2}\big)^{\frac12}$ & $\big(1-\frac{1}{ \xi^2}\big)^{\frac12}$\\ \hline
\end{tabular}
\caption{The coefficients $C, C'$ and $H$ in each order of expansion ($\xi=m l, \nu=-\mu l H$).}
\end{table}

If we write ADM metric then we will have the following functions according to  relation (\ref{ansatz}) 
\bea\label{polsol} 
T+X&=&2(C\,\rho^{\nu}+C_0)-\frac{2\rho}{l^2}\,\,\,,\quad \,\,T-X=2l^2(C\,\rho^{\nu}+C_0)+2\rho\,,\nn\\
Y&=&-2l(C\,\rho^{\nu}+C_0)\,\,\,,\quad \,\,R^2=-T^2+X^2+Y^2=\frac{4 \rho^2}{l^2}\,. 
\eea 
Note that the value of $C_0$ can not be fixed by equations of motion and it is a free parameter of solutions.
Using the above values we are able to compute the super-angular momentum.
By (\ref{JGR}) and (\ref{JEM}) we can find super-angular momentum  as
\be
\bs J=\frac{C_0}{\kappa l}\,C'\,\Big(l^2+1\,,-l^2+1\,,2l\Big)\,,
\ee
where $C'$ is given in the second column of table 2 which depends on the order of expansion of  Lagrangian. 
From  super-angular momentum and using relations in  (\ref{ang}) we can read angular momentum and mass. 

To do this we need to know the background solution and the value of $\Delta$. The background solution can be found by inserting $C_0=0$, so the value of super-angular momentum would be zero. On the other hand we suppose that $\Delta=0$ so we find 
\be
J=\frac{4\pi l C_0}{\kappa }C'\,,\quad\quad M=\frac{4\pi  C_0}{\kappa }C'\,.
\ee
But we need to show that the choice of $\Delta=0$ is consistent with the first law of thermodynamics for black holes. 
There is an alternative way to find the same value for mass. What we need is to check a consistency between the first law of  thermodynamics and Smarr-like formula $M=\frac12\,T_H S\,+\Omega_h J+\frac12\,\Phi_h \bar{Q}$. 

As we noted, the only free parameters of solutions are $Q$ and $C_0$. By differentiating Smarr-like formula with respect to these parameters we expect to find the first law. This fixes constant coefficients of Smarr-like formula and also gives the value of  mass. But to use Smarr-like formula we need to find temperature, angular velocity and the value of electric potential  at horizon, we also need the entropy of black hole. 

Horizon is a circle and it is parametrized by $\phi$. Location of this horizon is given by $N^2=\frac{R^2}{K^2}=0$. Since the radius of horizon  is given by $K$, we assume $K>0$. So the only value for the location of horizon will be $\rho=0$ if we have $\nu>0$.
Using (\ref{temp}) and (\ref{phih}) we find the following quantities at horizon
\bea 
\Omega_h=\frac{l(C\rho^\nu+C_0)}{l^2(C\rho^\nu+C_0)+\rho}\Big|_{\rho\rightarrow 0}=\frac{1}{l}\,,
\quad T_h=\frac{2\rho}{\pi l^2\sqrt{2\Big(l^2(C\rho^\nu+C_0)+\rho\Big)}}\Big|_{\rho\rightarrow 0}=0\,,
\quad \Phi_h=0\,.
\eea 
As we see the above quantities are independent of the order of expansion. The zero value of temperature indicates that we are dealing with an extremal black hole. The above values immediately give a relation between  mass and  angular momentum from  Smarr-like formula, which is $M=Jl$ and it shows that $\Delta=0$ is a correct assumption.

Finally entropy can be found by using Wald formula (\ref{ent1}) 
\be
S=\frac{2\pi\,A_h}{\kappa}\,C'\,,\quad\quad 
A_h=2\pi\sqrt{2\Big(l^2(C\rho^\nu+C_0)+\rho\Big)}\Big|_{\rho\rightarrow 0}=2\pi\sqrt{2l^2 C_0}\,.
\ee

As we told before, there is another solution when $\nu=0$. This is the well known BTZ solution in presence of a constant gauge field and has the same angular velocity, temperature, mass and entropy as nonzero solution.
\subsection{Logarithmic MCS-charged solution}
In addition to the self-dual solution of the previous subsection we can consider another ansatz which is also a solution of equations of motion. We call it  logarithmic MCS-charged solution \cite{Garbarz:2008qn}, \cite{Clement:2009ka}
\be \label{LMCS} 
{\bs X}={\bs \alpha}\,D \rho^\nu\ln{\rho}+{\bs \beta}\,\rho+{\bs \gamma} D_0\,,\quad{\cal A}=Q\,\rho^\frac{\nu}{2}\big(dt-l d\phi\big)\,. 
\ee 
Again $Q$ corresponds to the electric (magnetic) charge of solution. The equations of motion for $T, X$ or $Y$ restrict us to the following frame
\be \label{flog}
{\bs \alpha}={\bs \gamma}=\Big(1+l^2,1-l^2,-2l\Big)\,,\quad
{\bs \beta}=\Big(\frac{l^2-1}{l^2},-\frac{l^2+1}{l^2},0\Big)\,.
\ee 
Imposing the above frame in equations of motion for $T, X$ or $Y$, one finds the following relation for every level of expansion of Lagrangian
\be
c_1 \rho^\nu+c_2 \rho^\nu\ln\rho=0\,,
\ee
where values of $\nu$ and $D$ can be found by solving equations $c_1=c_2=0$. When order of expansion is greater than two then there will be two solutions for these equations, one solution is always $\nu=1$ and the other one can be written as a function, $\nu=\nu(m,l)$. 

Values of parameter $D$ at each level of expansion are given in table 3 by solving equations of motion for $X$, $Y$ or $T$. Equations of motion for gauge field components $A_1$ or $A_2$ give a relation which restricts the coupling constant of Maxwell-Chern-Simons $(\mu)$. The values of $\mu$ as a function of $m$ and $l$ are given in table 3. 

Again equation of motion for $\zeta$ gives values of  cosmological constant. These values at each level are the same as previous ones and are given in table 1. 
\begin{table}[ht] \renewcommand{\arraystretch}{1.5}
\center
\begin{tabular}{|c|c|c|c|}\hline
     & $\mu$ & $D$ & $D'$  \\ \hline 
${\mathcal{O}}(2)$ & $-\frac{1}{l}$ &$-\frac{\kappa Q^2}{4}$ & 1\\ \hline
${\mathcal{O}}(4)$ & $-\frac1l\,(1+\frac{1}{2 \xi^2})$ & $-\frac{\kappa Q^2}{4}\frac{2\xi^2+1}{2\xi^2-1}$ & $(1-\frac{1}{2 \xi^2})$ \\ \hline
${\mathcal{O}}(6)$ & $-\frac1l\,(1+\frac{1}{2\xi^2}+\frac{3}{8 \xi^4 })$ & $-\frac{\kappa Q^2}{4}\frac{8\xi^4+4\xi^2+3}{8\xi^4-4\xi^2-1}$ & $(1-\frac{1}{2 \xi^2}-\frac{1}{8 \xi^4 })$\\ \hline
${\mathcal{O}}(\infty)$& $-\frac{1}{l}\,(1-\frac{1}{\xi^2})^{-\frac12}$ & $-\frac{\kappa Q^2}{4}(1-\frac{1}{\xi^2})^{-1}$ & $(1-\frac{1}{\xi^2})^{\frac12}$\\ \hline
\end{tabular}
\caption{Constant parameters $\mu$, $D$ and $D'$  for logarithmic MCS black holes, ($\xi=m l$).}
\end{table}

By inserting the frame in (\ref{flog}) into solution (\ref{LMCS}), metric in its ADM form can be read as 
\bea\label{logsol}
T+X&=&2D\rho^\nu\ln{\rho}+2D_0-\frac{2\rho}{l^2}\,,\quad
T-X=2l^2 D\rho^\nu\ln{\rho}+2l^2 D_0+2\rho\,,\nn\\
Y&=&-2l(D \rho^\nu\ln{\rho}+D_0)\,,\quad
R^2=-T^2+X^2+Y^2=\frac{4 \rho^2}{l^2}\,. 
\eea 
In this solution $D_0$ is a free parameter and cannot be fixed by equations of motion.

For future proposes we need to read angular velocity, temperature and electric potential at horizon. We also need to know horizon's area. By the same argument as in previous ansatz, horizon will be located at $\rho=0$ if $\nu$ has a positive value.
Using equations (\ref{temp}) and (\ref{phih}) and by knowing  the location of horizon, we find the following values
\bea 
\Omega_h\!\!&=&\!\!\frac{l(D\rho^\nu\ln \rho+D_0)}{l^2(D\rho^\nu\ln\rho+D_0)+\rho}\Big|_{\rho\rightarrow 0}=\frac{1}{l}\,,
\quad T_h=\frac{2\rho}{\pi l^2\sqrt{2\Big(l^2(D\rho^\nu\ln\rho+D_0)+\rho\Big)}}\Big|_{\rho\rightarrow 0}=0\,,\nn\\
\Phi_h\!\!&=&\!\!0\,,\quad A_h=2\pi\sqrt{2\Big(l^2(D\rho^\nu\ln\rho+D_0)+\rho\Big)}\Big|_{\rho\rightarrow 0}=2\pi\sqrt{2l^2 D_0}\,.
\eea
\subsubsection{$\nu=1$}
In this case values of $\mu$ and $D$ are given in the first and in the second column of table 3. To find physical quantities of this black hole let us first compute super-angular momentum.
Using relations in equations (\ref{JGR}) and (\ref{JEM}) one finds 
\be\label{supnu1}
\bs J=\,\frac{D_0}{\kappa l}\,D'\Big(l^2+1,-l^2+1,2l\Big)\,,
\ee
where $D'$ is a constant and its values are given in the last column of table 3. Similar to the previous case here we can read angular momentum and  mass of the black holes from super-angular momentum. The background solution can be found by imposing $D_0=0$. Inserting $\Delta=0$ in (\ref{ang}) one finds the following values for angular momentum and mass 
\be \label{jmnu1}
J=\frac{4\pi l D_0}{\kappa}D'\,,\quad  M=\frac{4 \pi D_0}{\kappa}\,D'\,.
\ee
To check that $\Delta=0$ is a correct choice, we use Smarr-like formula $M=\frac12\,T_H S\,+\Omega_h J+\frac12\,\Phi_h \bar{Q}$. In this case the only free parameters are $Q$ and $D_0$. By differentiating Smarr-like formula with respect to these parameters we can find the first law of thermodynamics for black holes. Since we have an extremal black hole with vanishing temperature and since the value of electric potential at  horizon is zero we find again $M=J l$ and so $\Delta=0$. 

Using the Wald formula (\ref{ent1}) for entropy,  we have
\be\label{entnu1}
S=\frac{2\pi\,A_h}{\kappa}\,D'\,,\quad 
A_h=2\pi\sqrt{2\Big(l^2(D\rho^\nu\ln\rho+D_0)+\rho\Big)}\Big|_{\rho\rightarrow 0}=2\pi\sqrt{2l^2 D_0}\,.
\ee
\subsubsection{$\nu=\nu(m,l)$}
As noted before there is another solution for $\nu$ as a function of $m$ and $l$. This new value does not exist in ${\mathcal{O}}(2)$ but appears at higher orders. The values of $\nu$ are given in the first column of table 4. Similar to the previous case we can find values of $\mu$ and $D$, which are given in table 4.
\begin{table}[ht] \renewcommand{\arraystretch}{1.5}
\center
\begin{tabular}{|c|c|c|c|}\hline
     & $\nu$ & $\mu$ & $D$  \\ \hline 
${\mathcal{O}}(4)$ & $\frac12\pm\frac14 \Delta_4 $ & $\frac{2\pm\Delta_4}{4l}\,(1+\frac{1}{2\xi^2})$ & $\pm\frac{(6\xi^2+1)(2\xi^2+3\mp2\Delta_4)}{8{(2\xi^2-1)\Delta_4}}$ \\ \hline
${\mathcal{O}}(6)$ & $\frac12\,\pm\,\frac14 (\frac{\Delta_6}{2\xi^2+1})$ & $ \frac{(8\xi^4+4\xi^2+3)(2(2\xi^2+1)\,\pm\,\Delta_6)}{32l\xi^4(2\xi^2+1)}$ & $ \pm\frac{3(16\xi^6+16\xi^4+2\xi^2+1)(8\xi^4+12\xi^2+7\mp4\Delta_6)}{16(16\xi^6-6\xi^2-1)\Delta_6}$\\ \hline
${\mathcal{O}}(\infty)$& $\frac12\,\pm\,\frac{\xi}{2}$ & $\pm\,\frac{1}{2l}\,\frac{\xi(\xi\pm\,1)}{\sqrt{\xi^2-1}}$ & $\pm\frac38\,\frac{\xi(\xi\mp\,1)}{\xi\pm\,1}$\\ \hline
\end{tabular}
\caption{Constant parameters $\nu$, $\mu$ and $D$ for logarithmic MCS black holes, ($\xi=m l$). $\Delta_4=\sqrt{2+4\xi^2}$ and $\Delta_6=\sqrt{16\xi^6+16\xi^4+10\xi^2+3}$.}
\end{table}

Super-angular momentum and therefore values of angular momentum and mass can be found by using equations (\ref{supnu1}) and (\ref{jmnu1}). In fact the values of $D'$ in this case are exactly equal to values for $\nu=1$. So these solutions have the same angular momentum and mass.  On the other hand by computing  entropy from  Wald formula we find that these solutions have the same entropy (\ref{entnu1}), with similar values of $D'$  (table 3). These similarities come from similarity of asymptotic and  near horizon geometries of the two solutions, although the geometries between these two limits are different. 

Note that existence of horizon at $\rho=0$, restricts $\nu$ to $\nu>0$. This restricts the values of $\xi=ml$ in the first column of table 4. 
\subsection{M-charged solution}
Now let us turn off Chern-Simons term by inserting $\mu=0$. We call this solution Maxwell-charged or M-charged solution. We suppose the following ansatz
\be {\bs X}={\bs \alpha} B\,\rho^{\nu}\ln(\frac{\rho}{\rho_0})+{\bs \beta}\,\rho\,,\qquad{\cal A}=Q
\ln(\frac{\rho}{\rho_0})\,\big(dt-\omega
d\phi\big)\,,
\ee
where $Q$ again is electric (magnetic) charge and $B$ and $\rho_0$ are some constants. To solve equations of motion we consider the following frame which is more general than the previous frames   
\bea
{\bs \alpha}=\,\Big(\omega^2+1\,,-\omega^2+1\,,\,-2\omega\Big)\,,\quad
{\bs \beta}=\Big(\frac{l^2-1}{l^2}\,,-\frac{l^2+1}{l^2}\,,0\Big)\,.\quad
\eea
If we choose $\omega=l$ then we will obtain self-dual M-charged solutions with $E=F_{t \rho}=F_{\rho \phi}=B$.
Similar self-dual solutions have been studied in \cite{Clement:1995zt},\cite{Kamata:1995zu} and \cite{Moussa:1996gm}.

Using the above frame and ansatz and by inserting these into  $T, X$ or $Y$ equations of motion we end up with the following equation 
\be
c_1 \rho^\nu\log(\frac{\rho}{\rho_0})+c_2\rho^\nu+c_3=0\,,
\ee
where $c_1, c_2$ and $c_3$ are functions of $\nu, m$ and $l$ and according to the order of expansion they have different values.
The only consistent solution of this equation, independent of the order of expansion, will be the case $\nu=0$.
With this value, $c_1=0$ and the value of $B$ can be found from $c_2+c_3=0$. One can see values of $B$ in the second column of table 5 in each order of expansion. 
\begin{table}[t] \renewcommand{\arraystretch}{1.5}
\center
\begin{tabular}{|c|c|c|}\hline
     & $B$ & $B'$  \\ \hline 
${\mathcal{O}}(2)$ & 1 & 1 \\ \hline
${\mathcal{O}}(4)$ & $\frac{ 2\xi^2+1}{ 2\xi^2-1}$ & $\frac{ 4\xi^4 -24 \xi^2-5}{ 2\xi^2( 2\xi^2-1)}$ \\ \hline
${\mathcal{O}}(6)$ & $\frac{ 8\xi^4 +4\xi^2+3}{ 8\xi^4 -2\xi^2-1}$ & $\frac{64\xi^8-384\xi^6-384 \xi^4 -64 \xi^2-27}{ 8\xi^4 ( 8\xi^4 -4  \xi^2-1)}$\\ \hline
${\mathcal{O}}(\infty)$&$\frac{ \xi^2}{ \xi^2-1}$ & $\frac{\xi(\xi^2-7)}{(\xi^2-1)^\frac32}$ \\ \hline
\end{tabular}
\caption{Coefficients $B$ and $B'$ in each order of the expansion ($\xi=m l$).}
\end{table}

There are two other equations of motion for $A_1$ and $A_2$. These equations satisfy by choosing a proper ansatz for gauge field.
Equation of motion for $\zeta$ gives the values of cosmological constant $\Lambda$ which are exactly equal to previous results in table 1.

Considering all above values, the corresponding ADM metric contains the following functions
\bea \label{mch}
T+X&=&2\kappa B\,Q^2\ln(\frac{\rho}{\rho_0})-\frac{2\rho}{l^2}\,,\qquad T-X=2\kappa\omega^2 B\,Q^2\ln(\frac{\rho}{\rho_0})+2\rho\,,\nn\\
Y&=&-2\kappa\,B Q^2\omega\ln(\frac{\rho}{\rho_0})\,,\qquad R^2=-T^2+X^2+Y^2=-\frac{4 \rho}{l^2}(B \kappa Q^2 (l^2-\omega^2)\ln(\frac{\rho}{\rho_0})-\rho)\,. 
\eea
Using this solution we can compute super-angular momentum.
From (\ref{JGR}) and (\ref{JEM}) we find 
\be \label{mchsuper}
\bs J=\frac{ Q^2}{l}\,B'\Big(\omega^2+1\,,-\omega^2+1\,,\,2\omega\Big)\,\,,
\ee 
where $B'$ is  given in the second column of table 5. From super-angular momentum (\ref{mchsuper})
and using (\ref{ang}) we can read angular momentum and mass as follows  
\be
J=\frac{4\pi Q^2 \omega^2}{l}\,B'\,,\quad M=\frac{4\pi Q^2 \omega}{l}\,B' \,.
\ee
In this case similar to previous ones we choose $\Delta=0$ in (\ref{ang}). 

As we mentioned before, horizons are roots of equation $N^2=\frac{R^2}{K^2}=0$. So according to  metric (\ref{mch}) there are two horizons at $\rho=\rho_+$ and $\rho=\rho_-$ which are outer and inner horizons respectively ($\rho_+>\,\rho_-$). But $K^2$ changes sign for a certain value of $\rho=\rho_c$ such that for $\rho<\rho_c$ we encounter closed time-like curves \cite{Clement:1995zt}. 

In the extremal solution when we go to self-dual limit $|\omega|\rightarrow \pm l$, from (\ref{mch}) we can see that the location of horizon goes to $\rho=0$, which is not consistent with $\rho>\rho_c$. In fact this is a naked singularity which is located at infinite geodesic distance  \cite{Clement:1995zt}. Note that the extremal black holes such as those we described in previous subsections have a horizon at infinite proper distance \cite{Clement:1995zt}. Therefore we can not compute thermodynamical parameters of this self-dual solution because this is a horizon-less solution. 
\subsection{Geodesic Completeness}
In previous sections we found two types of black holes, the polynomial solution (\ref{polsol}) and the logarithmic solution (\ref{logsol}), where their horizon were located at $\rho=0$. At $\rho=0$  curvature scalars such as $R,R^{\mu\nu}R_{\mu\nu},R^{\mu\nu\rho\sigma}R_{\mu\nu\rho\sigma},...$ are finite. In addition each point outside the horizon is located at an infinite radial distance from the horizon due to  extremality of  black holes. We now find a condition that time-like geodesics approach to horizon in an infinite amount of time or in other words solutions become geodesically complete. 

The black hole solutions (\ref{polsol}) and  (\ref{logsol}) have two  Killing vectors $\prt_{t}$ and $\prt_{\phi}$ corresponding to two manifest symmetries of their metric. We define these two  Killing vectors as
\be 
\label{kvs} K^{\mu}=(\prt_{t})^{\mu}=(1,0,0),\quad L^{\mu}=(\prt_{\phi})^{\mu}=(0,0,1)\,.
\ee
Constants of motion for a geodesic can be expressed as
\be \label{com1} {\cal E}_{t}=-K_{\mu}\,\frac{d x^{\mu}}{d\tau}\,,\quad {\cal E}_{\phi}=L_{\mu}\,\frac{d x^{\mu}}{d\tau}\,,\ee
where $\tau$ is an affine parameter. In addition, we  have another constant of motion. Geodesic equation implies that the following quantity is constant along the path
\be\label{com2} 
\varepsilon=-g_{\mu\nu} \,\frac{d x^{\mu}}{d\tau}\,\frac{d x^{\nu}}{d\tau}\,,
\ee
where $\varepsilon=-1,\,0$ or $1$  for time-like, null or space-like geodesics respectively. Using constants in (\ref{com1}) and multiplying $g^{\rho\rho}$ on both sides of (\ref{com2}), the geodesic equation for metric (\ref{polsol}) and (\ref{logsol}) becomes
\bea \label{geoeq} 
&&\Big(\frac{d \rho}{d\tau}\Big)^2-2\big(C\,\rho^{\nu}+C_0\big)\,(l\,{\cal E}_{t}+{\cal E}_{\phi})-\frac{2\rho}{l^2}\,(l^2{\cal E}^2_{t}-{\cal E}^2_{\phi})+\frac{4 \rho^2}{l^2}\,\varepsilon=0\,,\nn\\
&&\Big(\frac{d \rho}{d\tau}\Big)^2-2\big(D\,\rho^{\nu}\,\ln{\rho}+D_0\big)\,(l\,{\cal E}_{t}+{\cal E}_{\phi})-\frac{2\rho}{l^2}\,(l^2{\cal E}^2_{t}-{\cal E}^2_{\phi})+\frac{4 \rho^2}{l^2}\,\varepsilon=0. 
\eea
When $\rho\rightarrow\,0$ geodesics will be regular  if ${\cal E}_{\phi}=-l\,{\cal E}_{t}$ (see for example \cite{Clement:1995zt} for a similar argument). Solving each equation in (\ref{geoeq}) for time-like geodesics ($\varepsilon=-1$) together with ${\cal E}_{\phi}=-l\,{\cal E}_{t}$ shows that, it takes an infinite amount of time to reach the horizon at $\rho=0$.
\section{Born-Infeld-TMG solutions}
In this section we extend our work and consider gravitational Chern-Simons action. We add topological Lagrangian (\ref{LTMG}) to  Born-Infeld  Lagrangian (\ref{BINMG}). Since gravitational Chern-Simons action has third order of derivative terms  we just consider ${\mathcal{O}}(2)$, ${\mathcal{O}}(4)$ and ${\mathcal{O}}(\infty)$ in our computations. We divide solutions into extremal and non-extremal black holes. 
\subsection{Extremal black holes}
In section 3 we found a number of self-dual solutions in presence of  the Maxwell-Chern-Simons term. Now we  add  gravitational Chern-Simons term to the Lagrangian. Since our ansatz in section 3 also works here and all steps are similar, we just write final results in their corresponding tables. 
\subsubsection{Polynomial charged solution}
A self-dual extremal charged black hole in presence of gravitational CS term (we call it  polynomial charged solution) is given by the following ansatz
\be 
\bs X=\bs \alpha C \rho^{\nu}+\bs \beta \rho+\bs \gamma C_0\,,\quad\mathcal A=Q \rho^{\frac{\nu}{2}}\,(dt-ld\phi)\,.
\ee
Angular momentum, mass and entropy of this black hole  are as follows
\be J=\frac{4\pi C_0 l}{\kappa}\,C'\,,\quad M=\frac{4\pi C_0}{\kappa}\,C'\,,\quad S=\frac{2\pi A_h}{\kappa}\,C'\,.
\ee
Together with the following physical quantities at horizon ($\rho_h=0$)
\be
A_h=2\pi\sqrt{2l^2 C_0}\,,\quad \Omega_h=\frac{1}{l}\,,\quad T_h=0\,,\quad \Phi_h=0\,.
\ee
All values of constant parameters $\nu$, $C$ and $C'$ are listed in table 6.
\begin{table}[ht]\renewcommand{\arraystretch}{1.5}
\center
\begin{tabular}{|c|c|c|c|}\hline
     & $\nu$ & $C$ & $C'$\\ \hline 
${\mathcal{O}}(2)$ & $-\mu\,l$ & $-\frac{\kappa Q^2\,\nu \eta}{4(\nu-1)}\,\frac{1}{(2\nu-1+\eta)}$ & $1-\frac{1}{\eta}$ \\ \hline
${\mathcal{O}}(4)$ & $-\mu\,l\big(1+\frac{1}{2\xi^2}\big)^{-1}$ & $-\frac{\kappa Q^2\,\nu \eta}{4(\nu-1)}\,\frac{\xi^2+\frac12+2\nu(\nu-1)}{\xi^2(2\nu-1)+\eta(\xi^2-\frac12-4\nu(\nu-1))}$ & $1-\frac{1}{2\xi^2}-\frac{1}{\eta}$ \\ \hline
${\mathcal{O}}(\infty)$& $-\mu\,l\sqrt{1-\frac{1}{\xi^2}}$ & $-\frac{\kappa Q^2\,\nu\eta}{4(\nu-1)}\frac{\xi^2-1+2\nu(\nu-1)}{[\xi^2-(2\nu-1)^2]\eta-(2\nu-1)\xi\sqrt{\xi^2-1}}(1-\frac{1}{\xi^2})^{-1}$ & $\sqrt{1-\frac{1}{\xi^2}}-\frac{1}{\eta}$ \\ \hline
\end{tabular}
\caption{$\xi=ml$ , $\mu_{G} l=\eta$}
\end{table}
\subsubsection{Logarithmic charged solutions} 
Another self-dual extremal charged black hole  is  logarithmic charged black hole. Solving equations of motion leads to
\be 
{\bs X}={\bs \alpha}\,D \rho^{\nu} \ln(\rho)+{\bs \beta}\,\rho+{\bs \gamma} D_0\,,\quad{\cal A}=Q
\rho^\frac{\nu}{2}\,\big(dt-l
d\phi\big)\,. 
\ee
Angular momentum, mass and entropy of this black hole are as follows 
\be J=\frac{4 \pi l D_0}{\kappa} D'\,,\quad M=\frac{4 \pi D_0}{\kappa} D'\,,\quad S=\frac{2 \pi A_h}{\kappa} D'\,.\ee
Together with the following values at horizon
\be 
A_h=2\pi\sqrt{2l^2 D_0}\,,\quad T_h=0 \,,\quad \Omega_h=\frac{1}{l} \,,\quad \Phi_h=0\,.
\ee
We have two sets of solutions in this case. The first case is $\nu=1$, see table 7, and the second case is   $\nu=\nu(m,l,\mu_G)$, see table 8. In both cases $\mu$ is given in the first column of table 7 and $D'$ in the last column of this table. The values of cosmological constant do not change and are given in table 1.
\begin{table}[ht]\renewcommand{\arraystretch}{1.5}
\center
\begin{tabular}{|c|c|c|c|}\hline
  $\nu=1$   & $\mu$ & $D$ & $D'$ \\ \hline 
${\mathcal{O}}(2)$ & $-\frac{\nu}{l}$ & $-\frac{\kappa Q^2}{4}\,\frac{1}{(1-\frac{1}{\eta})}$ & $1-\frac{1}{\eta}$   \\ \hline
${\mathcal{O}}(4)$ & $-\frac{\nu}{l}\,(1+\frac{1}{2\xi^2})$ & $-\frac{\kappa Q^2}{4}\,\frac{1+\frac{1}{2\xi^2}}{1-\frac{1}{2\xi^2}-\frac{1}{\eta}}$ & $1-\frac{1}{2\xi^2}-\frac{1}{\eta}$  \\ \hline
${\mathcal{O}}(\infty)$ & $-\frac{\nu}{l}\,(1-\frac{1}{\xi^2})^{-\frac12}$  & $-\frac{\kappa Q^2}{4}\,\Big[1-\frac{1}{\xi^2}-\frac{1}{\eta}\,\sqrt{1-\frac{1}{\xi^2}}\Big]^{-1}$ & $\sqrt{1-\frac{1}{\xi^2}}-\frac{1}{\eta}$\\ \hline
\end{tabular}
\caption{$\xi=ml$ , $\mu_{G} l=\eta$}
\end{table}
\begin{table}[ht]\renewcommand{\arraystretch}{1.5}
\center
\begin{tabular}{|c|c|c|}\hline
      & $\nu$ & $D$  \\ \hline 
${\mathcal{O}}(2)$ & $-\frac{\eta-1}{2}$ & $-\frac{\kappa\, Q^2\,\nu^2\eta}{4}\,\frac{1}{(2\nu-1)\eta+6\nu(\nu-1)+1}$  \\ \hline
${\mathcal{O}}(4)$ & $\frac12+\frac{1}{4\eta} \big(\xi^2\pm \sqrt{4\eta^2(\xi^2+\frac12)+\xi^4}\big)$ & $-\frac{\kappa\, Q^2\,\nu^2\eta}{4}\,\frac{\xi^2+\frac12+2\nu(\nu-1)}{(2\nu-1)[2\xi^2-1-16\nu(\nu-1)]\eta+(6\nu(\nu-1)+1)\,\xi^2}$ \\ \hline
${\mathcal{O}}(\infty)$& $\frac12+\frac{\xi\sqrt{\xi^2-1}}{4\eta}\,[1\pm \sqrt{1+\frac{4\eta^2}{\xi^2-1}}]$ & $-\frac{\kappa\, Q^2\,\nu^2\eta}{4}\,\frac{(\xi^2-1+2\nu(\nu-1))}{(2\nu-1)[\xi^2-1-8\nu(\nu-1)]\eta+(6\nu(\nu-1)+1)\xi \sqrt{\xi^2-1}}\,(1-\frac{1}{\xi^2})^{-1}$  \\ \hline
\end{tabular}
\caption{$\xi=ml$ , $\mu_{G} l=\eta$}
\end{table}
\subsection{Non-extremal black holes}
There are non-extremal black hole solutions for equations of motion. We have already found non-extremal warped $AdS_3$ black holes without the TMG term in \cite{Ghodsi:2010ev}. Similar to it we suppose the following ansatz
\be
\bs X=\bs \alpha \rho^2+\bs \beta \rho+\bs \gamma\,,\quad \mathcal A=Q\big(2zdt-(\rho+2\omega z)d\phi\big)\,.
\ee
To find a non-extremal warped solution to satisfy equations of motion for $T, X$ and $Y$ we need to choose the following frame 
\be {\bs \alpha}=\Big(\frac12,-\frac12,0\Big)\,,\quad{\bs \beta}=\Big(\omega,-\omega,-1\Big)\,,\quad
{\bs \gamma}=\Big(\frac{\beta^2\rho_0^2}{4z}+(1+\omega^2)z,-\frac{\beta^2\rho_0^2}{4z}+(1-\omega^2)z,-2z\omega\Big),
\ee
where $\beta^2=1-2z$. To have a solution free of  closed time-like curves we must restrict ourselves to $0<\beta^2<1$, \cite{Clement:2009gq}. One can obtain warped $AdS$ metric in ADM form 
\be ds^2=-\beta^2\,\frac{\rho^2-\rho_0^2}{r^2}\,dt^2+r^2\big(d\phi-\frac{\rho+(1-\beta^2)\,\omega}{r^2}\,dt\big)^2+\frac{1}{\beta^2 \zeta^2}\,\frac{d\rho^2}{\rho^2-\rho_0^2}\,,\ee
where $r^2=\rho^2+2\omega \rho+\omega^2 (1-\beta^2)+\frac{\beta^2 \rho_0^2}{1-\beta^2}$. 

After frame fixing each $T, X$ or $Y$ equations of motion equivalently give the same value for charge $Q$. The equations of motion for gauge field components also give another relation. At ${\mathcal{O}}(2)$  gauge field equations restrict $\mu=1$. But for other orders these equations give a relation between $z, m, \mu$ and $\mu_G$. Final results for charge $Q$ and  value of $z$ are written in first two columns of  table 9. 

The equation of motion for $\zeta$ gives  cosmological constant in terms of other parameters of theory (see the last column of the table 9).
\begin{table}[ht]\renewcommand{\arraystretch}{1.5}
\center
\begin{tabular}{|c|c|c|c|}\hline
     & $Q^2$ & $z$ & $\Lambda$  \\ \hline 
$\!\!\!{\mathcal{O}}(2)\!\!\!$ &\!\!\! $\frac{3-2\mu_{G}}{2\kappa\,\mu_{G}}$ \!\!\!& \!\!\!$ $\!\!\! & $\!\!\!-\frac14+\frac{2\mu_G-1}{2\mu_{G}}z\!\!\!$\\ \hline
$\!\!\!{\mathcal{O}}(4)\!\!\!$ &\!\!\! $\frac{(\frac32+(\mu-2)\mu_G)m^2+2\mu_G}{\kappa \mu_G(\mu m^2+1)}\!\!\!$ & \!\!\!$\frac{2(\mu m^2+1)((\mu-1) m^2-\frac18)\mu_{G}}{((\mu+2)\mu_G-\frac32) m^2}$\!\!\! & $\!\!\!\frac{( 48{\kappa}^{2}{Q}^{4}-128\kappa{Q}^{2}+64) {z}^{2}+ ( 96- (64{m}^{2}+56) \kappa{Q}^{2}) z-1-16{m}^{2}} {64m^2}+\frac{z}{\mu_G}\!\!\!$ \\ \hline
$\!\!\!{\mathcal{O}}(\infty)\!\!\!$& $\!\!\!\frac{4(3\mu-2\mu_G)m^2-3\mu+18\mu_G}{2\kappa\mu^2\mu_G(4m^2+3)}\!\!\!$ & \!\!\!$\frac{-2(m^2+\frac34)((\mu^2-1)m^2-\frac14\mu^2)\mu_G}{3m^2(\mu-2\mu_G)}$\!\!\! & $\!\!\!-\frac{\mu}{3}+\frac23\frac{(2\mu-1)(\mu-1)}{\mu}\!\!\!$\\ \hline
\end{tabular}
\caption{Parameters given by equations of motion for warped $AdS_3$ solution.}
\end{table}

From equations (\ref{JGR}) and (\ref{JEM}) we  find the following value for super-angular momentum 
\be
\bs J=\Xi_1 \rho_0^2\big(-1,1,0\big) + \Xi_2 \big(\omega^2+1,-\omega^2+1,2\omega\big)\,,
\ee
where coefficients are given in table 10, (
note that the value of $z$ at each row must be read from second column of table 9).
\begin{table}[ht]\renewcommand{\arraystretch}{1.5}
\center
\begin{tabular}{|c|c|c|}\hline
      & $\Xi_1$ & $\Xi_2$  \\ \hline 
$\!\!\!{\mathcal{O}}(2)\!\!\!$ & $\frac{(z-\frac12)}{8\kappa\,z\,\mu_{G}}(4z-2\mu_{G}-1)$ &  $-\frac{z(z-\frac12)}{2\kappa\mu_{G}} (2\mu_{G}-1)$ \\ \hline
$\!\!\!{\mathcal{O}}(4)\!\!\!$ & $\frac{(z-\frac12)}{8\kappa\,z\,\mu_{G}}(4z-2\mu_G(2-\mu)-1)$ &  $-\frac{z(z-\frac12)}{2\kappa\mu_G}(2\mu_G(2-\mu)-1)$ \\ \hline
$\!\!\!{\mathcal{O}}(\infty)\!\!\!$ & $\frac{(z-\frac12)}{8\kappa\,z\,\mu\mu_{G}}((4z-1)\mu-2\mu_G)$ &  $-\frac{z(z-\frac12)}{2\kappa\mu\mu_G}(2\mu_G-\mu)$ \\ \hline
\end{tabular}
\caption{Coefficients corresponding to super-angular momentum for warped $AdS_3$ solution.}
\end{table}

To find angular momentum and mass we must subtract the values of background. Background is given by inserting $\rho_0=\omega=0$. Using equation (\ref{ang}) we can read angular momentum and mass. For non-extremal black holes we consider $M=4\pi(\delta{\bs J}^Y)$ which we show to be consistent with the first law of thermodynamics for black holes. After subtraction we find 
\be\label{nonjm}
J=4\pi(-\Xi_1\rho_0^2+\Xi_2\omega^2)\,,\quad M=8\pi\Xi_2\omega\,.
\ee
The location of the horizon is given by $\rho=\rho_0$ or equivalently $r_h=\frac{\rho_0+2\omega z}{\sqrt{2z}}$.
Quantities such as area of horizon, Hawking temperature, angular velocity and electric potential at horizon can be found by using relations in (\ref{temp}) and (\ref{phih})
\be 
A_h=2\pi r_h=\frac{2\pi}{\sqrt{2z}}\,(\rho_0+2\omega\,z)\,,\quad T_H=\frac{(1-2z)\,\rho_0}{A_h}\,,\quad\Omega_h=\frac{2\pi\sqrt{2z}}{A_h}\,,\quad \Phi_h=0\,.
\ee
Again entropy can be found by adding Wald entropy and contribution from  Chern-Simons term (\ref{ent2})
\be
S=\frac{16\pi^2}{(1-2z) \sqrt{2z}}(2z\Xi_1\rho_0+\Xi_2\omega)\,.
\ee
Using Smarr-like formula $M=T_H S\,+2\Omega_h J+\frac12\,\Phi_h \bar{Q}$ we can check that thermodynamical quantities satisfy the first law of black hole thermodynamics, $dM=T_H dS+\Omega_h dJ$. To do this we consider $M=M(\rho_0,\omega), S=S(\rho_0,\omega)$ and $J=J(\rho_0,\omega)$ then differentiate with respect to $\rho_0$ and $\omega$. The value of mass from Smarr-like formula agrees exactly with $M=4\pi(\delta{\bs J}^Y)$. This value of mass is exactly equal to the ADT mass of the black hole.
\section{Central charges of  dual CFTs, Cardy's formula approach}
According to  AdS/CFT conjecture \cite{Maldacena:1997re} one may expect that for some sectors of three dimensional gravities which are either asymptotically $AdS_3$ or $AdS_3$-like, there exists a  two dimensional dual conformal field theory. In this work we found such sectors for different types of massive gravities which were asymptotically $AdS_3$ (extremal black holes) or $AdS_3$-like (non-extremal warped-$AdS_3$ black holes), so it will be interesting to find some properties of these two dimensional dual conformal field theories. 

$AdS_3$ metric in Poincar$\grave{e}$ coordinates is given by
\be \label{ads} 
ds^2=l^2\,\frac{dx^{+}\,dx^{-}+dy^2}{y^2}\,,
\ee
where $l$ is the length of $AdS_3$ and boundary is located at $y=0$. 
The global symmetry of $AdS_3$ is $SO(2,2)$. 
The $SO(2,2)$ algebra, gets enhanced to asymptotic isometery algebra, which coincides with two copies of the Witt algebra.

The algebra of asymptotical conserved charges associated to asymptotic  Killing vectors satisfy two copies of  Virasoro algebra.
This algebra has equal left and right central charges  $c_L=c_R=\frac{3l}{2G}$  \cite{BH}, where  $G$ is the three dimensional gravitational constant. 
It is believed that these are central charges of a  two dimensional conformal field theory living on the boundary of $AdS_3$. 

In this paper we consider gravities which have been shown to be unitary in bulk and  boundary \cite{Gullu:2010em}. This property allows us to use Cardy's formula to read the central charges of  the dual CFT theories in terms of black hole entropies, i.e.
\be 
\label{card1} S=\frac{\pi^2}{3}\big(c_L\,T_L+ c_R\,T_R\big)\,,
\ee 
where $T_L$ and $T_R$ are left and right temperatures. Alternatively one may use the following relation 
\be \label{card2} 
S=2\pi \big(\sqrt{\frac{c_L\,E_L}{6}}+\sqrt{\frac{c_R\,E_R}{6}}\,\big)\,,
\ee
where $E_{L}$ and $E_{R}$ are the left and right energies and depending on each solution they have different values.

In \cite{Kraus:2005vz} it has been shown that for asymptotic $AdS_3$ sector of pure gravity we can find a CFT dual with a central function 
$ c=\frac{l}{2G}\,g_{\mu\nu}\,\frac{\prt \mathcal L_3}{\prt R_{\mu\nu}}$.
In the following we will show that this relation only works for those sectors which are asymptotically $AdS_3$ and does not give a correct result for asymptotically warped-$AdS_3$ sectors.
\subsection{Asymptotic $AdS_3$ sectors}
In previous sections we found two sets of extremal solutions, logarithmic solution and  polynomial solution with(without) presence of gravitational Chern-Simons term. These solutions have the following metrics 
\bea \label{esol}
ds_{log}^2&=&(2D\rho^\nu\ln{\rho}+2D_0-\frac{2\rho}{l^2}) dt^2+(2l^2 D\rho^\nu\ln{\rho}+2l^2 D_0+2\rho) d\phi^2-4l(D \rho^\nu\ln{\rho}+D_0)dt d\phi+\frac{l^2 d\rho^2}{4 \rho^2} \,,\nn \\
ds_{poly}^2&=&(2C\rho^\nu+2C_0-\frac{2\rho}{l^2}) dt^2+(2l^2 C\rho^\nu+2l^2 C_0+2\rho) d\phi^2-4l(C \rho^\nu+C_0)dt d\phi+\frac{l^2 d\rho^2}{4 \rho^2} \,.
\eea
In both cases Hawking temperature is zero and we have extremality condition $J=Ml$. As we told before, the existence of a horizon is possible if $\nu>0$.
Asymptotic behavior can be found by sending $\rho$ to infinity. This requires $\nu<1$ to have an asymptotically $AdS_3$ solution.

Since both these solutions have the same asymptotic symmetry and  belong to the asymptotic $AdS_3$ sector of three dimensional massive gravites, we expect that both have the same two dimensional dual CFTs. The central charges of these dual CFTs can be obtained from (\ref{card2}) by defining left and right energies as a linear combination of  mass and angular momentum. We define
\be\label{ELR}
E_{L}\equiv \frac{M l+ J}{2}\,,\quad  E_{R}\equiv \frac{M l- J}{2}\,.
\ee
By knowing entropy we can read the left central charge of the dual CFT. We have done this and all results for  $c_L$  are listed in table 11. As we see these central charges include corrections to $\frac{3l}{2G}$ found in  \cite{BH}. These corrections are coming from higher curvature terms and gravitational Chern-Simons term. 
\begin{table}[ht] \renewcommand{\arraystretch}{1.5}
\center
\begin{tabular}{|c|c|c|}\hline
     & $c_L$ & $c_R$ \\  \hline 
${\mathcal{O}}(2)$ & $\frac{3 l}{2 G}(1-\frac{1}{\eta})$ & $\frac{3 l}{2 G}(1+\frac{1}{\eta})$ \\ \hline
${\mathcal{O}}(4)$ & $\frac{3 l}{2 G}\,(1-\frac{1}{2 \xi^2}-\frac{1}{\eta})$ & $\frac{3 l}{2 G}\,(1-\frac{1}{2 \xi^2}+\frac{1}{\eta})$ \\ \hline
${\mathcal{O}}(6)$ & $\frac{3 l}{2 G}\,(1-\frac{1}{2 \xi^2}-\frac{1}{8 \xi^4}-\frac{1}{\eta})$ & $\frac{3 l}{2 G}\,(1-\frac{1}{2 \xi^2}-\frac{1}{8 \xi^4}+\frac{1}{\eta})$ \\ \hline
${\mathcal{O}}(\infty)$ & $\frac{3 l}{2 G}\,((1-\frac{1}{\xi^2})^\frac12-\frac{1}{\eta})$ & $\frac{3 l}{2 G}\,((1-\frac{1}{\xi^2})^\frac12+\frac{1}{\eta})$ \\ \hline
\end{tabular}
\caption{Central charges of dual CFTs}
\end{table}

Let us review the most important results:

1. Because of definition of  left and right energies (\ref{ELR}) we can find just the left central charge of the dual CFT. 

2. Since both logarithmic or polynomially solutions are asymptotically similar and only have different fluctuations around the $AdS_3$ space-time we expect to find the same central charges at each level of expansion.

3. The central charges without considering the gravitational Chern-Simons term can be obtained by sending the parameter $\eta=\mu_G l\rightarrow\infty$.

4. The value of the left central charge at each level of  expansion can be found by expanding the ${\mathcal{O}}(\infty)$ result with respect to $\xi=m l$.
\subsection{Asymptotic warped-$AdS_3$ sectors}
In non-extremal warped-$AdS$ solution the $SL(2,R)_L\times SL(2,R)_R$ symmetry breaks into the isometery group of $SL(2,R)\times U(1)$ \cite{Anninos:2008fx}. So this solution belongs to a different sector of three dimensional massive gravities and we expect different values for central charges of  dual CFTs. In this case we find both  left and right central charges by using Cardy's formula. 

We can either use the Cardy's formula in (\ref{card1}), where  left and  right temperatures $T_{L/R}$  can be defined as 
\be 
T_L\equiv\frac{1-2z}{2\pi\sqrt{2z}}\rho_0\,,\qquad T_R\equiv\frac{1-2z}{2\pi\sqrt{2z}} 2\omega z\,.
\ee
or we can use (\ref{card2}) with the following left and right energies \footnote{We can also use  $E_{L}=\frac{\pi^2}{6}\,c_{L}T^2_{L}$ and $E_{R}=\frac{\pi^2}{6}\,c_{R}T^2_{R}$.}
\be
E_L=\frac12\omega M-J=4\pi\Xi_1\rho_0^2\,,\qquad E_R=\frac12 \omega M=4\pi \Xi_2\omega^2\,.
\ee 
Both ways give the following values for  left and right central charges
\be
c_L=3.2^6\pi\frac{z}{(1-2z)^2}\Xi_1\,,\qquad c_R=3.2^4\pi\frac{1}{z(1-2z)^2}\Xi_2\,,
\ee
where values for $\Xi_1$ and $\Xi_2$ are given in table 10 and values for $z$ are listed in table 9. 

Using the above values for left and right central charges we can find holographic gravitational anomaly. Using parameters in table 10 we see that the value of anomaly is independent of  order of expansion
\be
c_L-c_R=\frac{3.2^4\pi}{(1-2z)^2}(4z\Xi_1-\frac{1}{z}\Xi_2)=\frac{3}{G\mu_G}\,.
\ee
This value for anomaly agrees with the value in \cite{Kraus:2005zm} exactly . 

There are important points to note:

1. The difference between values of  left and right central charges only depends on  gravitational Chern-Simons coupling. 

2. Unlike the asymptotic $AdS$ sector here  central charges of dual CFTs are independent and can not be  obtained from  expansion of the ${\mathcal{O}}(\infty)$ result.

Although in this section we found various results of central charges for dual CFTs we must check them from a more accurate approach. We will show that computing the asymptotic conserved charges allows us to find  central charges and confirms our computations in this section.
\section{Asymptotic conserved charges for extremal solutions}
As we saw in the previous section, Cardy's formula just gives left central charges of CFTs dual to asymptotically $AdS_3$ sectors. In this section we try to use another approach to find central charges by using asymptotic properties of  solutions. We will show that this approach gives both left and right central charges and confirms  results of Cardy's formula.

To compute conserved charges such as mass and  angular momentum associated to  Killing vectors of a typical background, we must linearize equations of motion around this background. According to Abbott-\,Deser (AD) formalism \cite{Abbott:1981ff} these conserved charges are expressed as 
\be 
Q^{\mu}(\bar\xi)=\frac{1}{8\pi G}\int_{\mathcal M}d^{D-1}x\,\sqrt{-\bar g}\,\bar\xi_{\nu}\,\delta T^{\mu\nu}\,,
\ee
where $\delta T^{\mu\nu}$ is the linearized energy-momentum tensor and $\bar\xi_{\nu}$ is a background  Killing vector. The value of $\bar\xi_{\nu}\delta T^{\mu\nu}$ generates a conserved current whose spatial integral for different components gives  conserved charges. 
Computations of conserved charges in presence of higher curvature gravity theories has been done firstly  for $AdS$ background in \cite{Deser:2002jk}. For an arbitrary background, calculations have been discussed in \cite{Bouchareb:2007yx}. 

The  Killing vectors $\bar\xi_{\nu}$ are  generators of isometeries of background metric, but we also have asymptotic  Killing vectors $\zeta_{\mu}$ which are defined as generators of non-trivial charges
\be 
\delta_{\zeta}h_{\mu\nu}=\nabla_{\mu} \zeta_{\nu}+\nabla_{\nu} \zeta_{\mu}.
\ee
Existence of such asymptotic Killing vectors is due to the fact that  killing equations do not fall-off fast enough near boundary. For any consistent set of boundary conditions one can find an associated asymptotic symmetry group (ASG), which is defined as a set of symmetry transformations modulo the set of trivial symmetry transformations\cite{Guica:2008mu}. 
\subsection{Linearization of equation of motion and conserved currents}
\subsubsection*{$\bullet$ Curvature terms}
To find  conserved charges, AD formalism suggests  linearization of equations of motion. Let us start with  pure gravity terms in Lagrangian up to  cubic terms of  curvature (\ref{EXL})
\bea 
\label{Lg} \mathcal{L}_3&=&\sqrt{-g}\,\, \Big\{R-2\Lambda+\kappa_1\,R^2+\kappa_2\,R_{\mu\nu}R^{\mu\nu}+\kappa_3\,R^{\mu}_{\,\nu}R^{\nu\rho}R_{\rho\mu}+\kappa_4\,R\, R_{\mu\nu}R^{\mu\nu}+ \kappa_5 \,R^3\Big\}\,,\nn\\
\kappa_1&=&\frac{3}{8m^2}\,,\quad \kappa_2=-\frac{1}{m^2}\,,\quad \kappa_3=-\frac{2}{3m^4}\,,\quad \kappa_4=\frac{3}{4m^4}\,,\quad \kappa_5=-\frac{17}{96m^4}\,.
\eea 
Equations of motion for Lagrangian (\ref{Lg}) are given by
\bea\label{EOM} 
T_{\mu\nu}&=&R_{\mu\nu}-\frac12 g_{\mu\nu} R+\Lambda g_{\mu\nu}+2\kappa_1  R (R_{\mu\nu}-\frac14 g_{\mu\nu} R)+(2\kappa_1+\kappa_2) (g_{\mu\nu}\Box-\nabla_{\mu}\nabla_{\nu})R+\kappa_2  \Box ( R_{\mu\nu}-\frac12g_{\mu\nu} R)\nn\\
&+&2\kappa_2 (R_{\mu\rho\nu\sigma}-\frac14 g_{\mu\nu} R_{\rho\sigma}) R^{\rho\sigma}+\kappa_3  \Big(3 R_{\mu\alpha}R^{\alpha\beta}R_{\beta\nu}+\frac32 \big[ g_{\mu\nu} \nabla_{\alpha}\nabla_{\beta} R^{\alpha\rho}R_{\rho}^{\beta}+\Box R_{\mu}^{\alpha}R_{\alpha\nu}-2 \nabla_{\alpha}\nabla_{(\mu}R_{\nu)}^{\beta}R_{\beta}^{\alpha} \big]\nn\\
&-&\frac12 g_{\mu\nu} R^{\alpha}_{   \beta}R^{\beta\rho}R_{\rho\alpha}\Big)
+\kappa_4  \Big(R_{\mu\nu} R_{\alpha\beta}R^{\alpha\beta}+2R  R_{\mu}^{\alpha}R_{\alpha\nu}+ g_{\mu\nu} \nabla_{\alpha}\nabla_{\beta} R^{\alpha\beta}R+\Box R R_{\mu\nu}-2 \nabla_{\alpha}\nabla_{(\mu}R_{\nu)}^{\alpha} R\nn\\
&-&[\nabla_{\mu}\nabla_{\nu}-g_{\mu\nu} \Box ] (R_{\alpha\beta}R^{\alpha\beta})-\frac12 g_{\mu\nu}R R_{\alpha\beta}R^{\alpha\beta}\Big)+\kappa_5\Big(3 R_{\mu\nu}R^2+3 [ g_{\mu\nu}\Box-\nabla_{\mu}\nabla_{\nu}] R^2-\frac12 g_{\mu\nu}R^3\Big) .
\eea
For AdS background in $D$ dimensions we have the following relations
\be
\label{adsback} R_{\mu\alpha\nu\beta}=\frac{2\Lambda_0}{(D-1)(D-2)}(g_{\mu\nu} g_{\alpha\beta}-g_{\mu\beta}g_{\nu\alpha})\,,\quad R_{\mu\nu}=\frac{2\Lambda_0}{(D-2)}g_{\mu\nu}\,,\quad R=\frac{2\Lambda_0 D}{(D-2)}\,,
\ee
where $\Lambda_0$ is a proper cosmological constant. If we insert background (\ref{adsback}) into equations of motion (\ref{EOM}) then we will find a critical value for $\Lambda$ in terms of $\Lambda_0$ 
\be \label{eqlam} 
\Lambda=\Lambda_0+2\Lambda_0^2(D\kappa_1 +\kappa_2)\frac{D-4}{(D-2)^2}+4\Lambda_0^3(\kappa_3+D \kappa_4+D^2 \kappa_5)\frac{D-6}{(D-2)^3}\,.
\ee
We now suppose small fluctuations around asymptotic background metric ($AdS$ space) as $g_{\mu\nu}=\bar g_{\mu\nu}+h_{\mu\nu}$ and use it to linearize equations of motion. Defining 
\be\mathcal{G}_{\mu\nu}\equiv R_{\mu\nu}-\frac12\,g_{\mu\nu} R+\Lambda g_{\mu\nu}\,,\ee
one finds the following linearized parts ($h\equiv\bar g^{\mu\nu}\,h_{\mu\nu}$)
\bea 
\delta\mathcal{G}_{\mu\nu}&=& \delta R_{\mu\nu}-\frac12\,\bar g_{\mu\nu} \delta R-\frac{2\Lambda_0}{D-2} \,h_{\mu\nu}\,,\nn\\
\delta R_{\mu\nu}&=&\frac12\,(\bar\nabla^{\lambda}\bar\nabla_{\mu}h_{\lambda\nu}+\bar\nabla^{\lambda}\bar\nabla_{\nu}h_{\lambda\mu}-\bar\Box h_{\mu\nu}-\bar\nabla_{\mu}\bar\nabla_{\nu}h)\,,\nn\\
\delta R&=&(\bar\nabla^{\lambda}\bar\nabla^{\rho}h_{\lambda\rho}-\bar\Box h\,)-\frac{2\Lambda_0}{D-2}\,h\,.
\eea 
Linearization of equation of motion (\ref{EOM}) is lengthy so we have summarized details of calculations in Appendix A. Using relations in (\ref{appA1}) and (\ref{appA2}) we find ($\Upsilon\equiv\frac{2\Lambda_0}{D-2}$)
\bea \label{LEOM}  
\delta T_{\mu\nu}&=&\Big(1+2(D \kappa_1+\kappa_2) \Upsilon+( 9 \kappa_3+5D \kappa_4+3D^2 \kappa_5 ) \Upsilon^2 \Big) \delta\mathcal{G}_{\mu\nu}+\kappa_2  \Big(\bar\Box \delta\mathcal{G}_{\mu\nu}-\frac{D-2}{D-1} \Upsilon \bar g_{\mu\nu}\delta R\Big)\nn\\
&+&\Big(2\kappa_1+\kappa_1+\big(3\kappa_3+(D+4) \kappa_4+6D \kappa_5\big) \Upsilon\Big) \Big( \bar g_{\mu\nu} \bar\Box-\bar\nabla_{\mu}\bar\nabla_{\nu}+\Upsilon \bar g_{\mu\nu}\Big)\delta R\nn\\
&+&\Big(3\kappa_3+D \kappa_4\Big) \Upsilon \Big(\bar\Box \delta\mathcal{G}_{\mu\nu}-2\bar\nabla^{\alpha}\bar\nabla_{(\mu}\delta\mathcal{G}_{\nu)\alpha}+\bar g_{\mu\nu} \bar\nabla^{\alpha}\bar\nabla^{\beta}\delta\mathcal{G}_{\alpha\beta}\Big)\nn\\
&+&\Big(\Lambda-\frac{(D-2)}{2} \Upsilon-\frac{(D-4)}{2} \Upsilon^2 (D \kappa_1+\kappa_2)-\frac{(D-6)}{2} \Upsilon^3 (\kappa_3+D \kappa_4+D^2 \kappa_5)\Big) h_{\mu\nu} .
\eea
The last line vanishes by (\ref{eqlam}). We can also use relation $\bar\nabla^{\beta}\delta\mathcal{G}_{\alpha\beta}=0$ to simplify the above equation.

Asymptotic conserved charges are given by the following integration
\be 
\label{cc1} Q(\bar{\xi})=\frac{1}{8\pi\,G}\int_{\mathcal M} d^{D-1}x \sqrt{-\bar g}\, \mathcal {K}^0\,,
\ee
where $\mathcal K^{\mu}=\bar \xi_{\nu}T^{\mu\nu} $ and $\mathcal M$ is a spatial $D-1$ dimensional manifold.
To find conserved charges we should use  Killing equations for asymptotic Killing vectors $\bar{\xi}_{\mu}$ which satisfy  Killing vector equation $\bar\nabla_{\mu}\,\bar{\xi}_{\nu}+\bar\nabla_{\nu}\,\bar{\xi}_{\mu}=O(h).$ Since we have already linearized equations to $O(h)$ we can ignore $O(h)$ due to this  Killing equation and put it to zero. We also use the following relations to simplify our results
\be \label{app1}
\bar\nabla^{\alpha}\bar\nabla_{\beta}\bar \xi_{\nu}={\bar R^{\mu}}_{\nu\alpha\beta} \bar \xi_{\mu}\,,\quad \bar\Box\bar \xi_{\mu}=-\Upsilon \bar \xi_{\mu}\,,\quad \bar \xi_{\nu}\,\bar\nabla_{\alpha}\bar\nabla^{\mu}\mathcal{G}^{\alpha\nu}=\frac{\Upsilon}{D-1}(D\bar \xi_{\nu}\delta\mathcal{G}^{\mu\nu}-\bar \xi^{\mu}\delta{\mathcal{G}^{\alpha}_{\,\alpha}})\,,\quad \delta{\mathcal{G}^{\alpha}_{\,\alpha}}=-\frac{\Lambda}{\Upsilon}\delta R\,.
\ee
Finally one finds the following conserved current
\bea \mathcal 
K^{\mu}&=&\Big(1+2(D \kappa_1+\kappa_2)\,\Upsilon+3\,(\kappa_3+D \kappa_4+D^2 \kappa_5)\,\Upsilon^2\Big)\bar \xi_{\nu}\delta\mathcal{G}^{\mu\nu}\nn\\
&+&\Big(2\kappa_1+\kappa_2+\big(3\kappa_3+(D+4)\kappa_4+6D \kappa_5 \big)\,\Upsilon\Big)\,\bar\nabla_{\alpha}\big\{\bar\xi^{\mu}\,\bar\nabla^{\alpha} \delta R-\bar\xi^{\alpha}\bar\nabla^{\mu} \delta R+\delta R \bar\nabla^{\mu}\bar\xi^{\alpha}\big\}\nn\\
&+&\Big(\kappa_2+(3\kappa_3+D \kappa_4)\,\Upsilon\Big)\,\bar\nabla_{\alpha}\big\{\bar \xi_{\nu}\bar\nabla^{\alpha} \delta\mathcal G^{\mu\nu}-\bar \xi_{\nu}\bar\nabla^{\mu} \delta\mathcal G^{\alpha\nu}-\delta\mathcal G^{\mu\nu}\bar\nabla^{\alpha}\bar \xi_{\nu}+ \delta\mathcal G^{\alpha\nu}\bar\nabla^{\mu}\bar \xi_{\nu}\big\}\,.
\eea
In first term, $\bar \xi_{\nu}\,\delta\mathcal{G}^{\mu\nu}$ can be written as a total derivative term, so the above expression will be a total derivative
\bea \label{current1} 
\mathcal{K}^{\mu}&=&\bar\nabla_{\nu}\,\Big\{\Big(1+2(D \kappa_1+\kappa_2)\,\Upsilon+3\,(\kappa_3+D \kappa_4+D^2 \kappa_5)\,\Upsilon^2\Big){\cal F}_1^{\mu\nu}\nn\\
&+&\Big(2\kappa_1+\kappa_2+\big(3\kappa_3+(D+4)\kappa_4+6D\kappa_5\big)\,\Upsilon\Big)\,{\cal F}_2^{\mu\nu}+\Big(\kappa_2+(3\kappa_3+D \kappa_4)\,\Upsilon\Big)\,{\cal F}_3^{\mu\nu}\Big\}\,,\nn\\
{\cal F}_1^{\mu\nu}&=&\frac12\big(\xi_{\rho}\bar\nabla^{\mu}h^{\nu\rho}+\xi^{\mu}\bar\nabla^{\nu}h+h^{\mu\rho}\bar\nabla^{\nu}\xi_{\rho}+\xi^{\nu}\bar\nabla_{\rho}h^{\mu\rho}+\frac12 h\bar\nabla^{\mu}\xi^{\nu}\big)-(\mu\leftrightarrow\nu),\nn\\
{\cal F}_2^{\mu\nu}&=&\bar\xi^{\mu}\,\bar\nabla^{\nu} \delta R-\bar\xi^{\nu}\,\bar\nabla^{\mu} \delta R+\delta R\bar\nabla^{\mu}\bar\xi^{\nu}\nn\\
{\cal F}_3^{\mu\nu}&=&\bar \xi_{\alpha}\bar\nabla^{\nu} \delta\mathcal G^{\mu\alpha}-\bar \xi_{\alpha}\bar\nabla^{\mu} \delta\mathcal G^{\nu\alpha}-\delta\mathcal G^{\mu\alpha}\bar\nabla^{\nu}\bar \xi_{\alpha}+ \delta\mathcal G^{\nu\alpha}\bar\nabla^{\mu}\bar \xi_{\alpha}\,.
\eea

\subsubsection*{$\bullet$ Gravitational Chern-Simons term}
Equation of motion for gravitational Chern-Simons Lagrangian (\ref{LTMG}) is
\be
\label{eomTMG} T_{\mu\nu}=\frac{1}{\mu_{G}}\,C_{\mu\nu}=\frac{1}{\mu_{G}}\varepsilon_{\mu}^{\,\,\alpha\beta}\,\nabla_{\alpha}(R_{\beta\nu}-\frac14\,g_{\beta\nu}\,R)\,,
\ee
where $\varepsilon_{\mu\nu\rho}=\sqrt{-g}\epsilon_{\mu\nu\rho}$ with $\epsilon_{012}=-1$. The linearized form of (\ref{eomTMG}) is \cite{Bouchareb:2007yx, Liu:2009pha }
\be 
\delta C_{\mu\nu}=\varepsilon_{\mu}^{\,\,\alpha\beta}\,\bar\nabla_{\alpha}(\delta R_{\beta\nu}-\frac14 
\bar g_{\beta\nu}\,\delta R-\Upsilon\,h_{\beta\nu})\,,
\ee
and conserved current constructed out of this term can be expressed as 
\be 
\mathcal K^{\mu}=\frac{1}{\mu_{G}}\bar\xi_{\nu} \delta C^{\mu\nu}=\frac{1}{2\mu_{G}} \bar\nabla_{\alpha} \big\{\varepsilon^{\mu\alpha\beta}\delta\mathcal G_{\,\nu\beta} \bar\xi^{\nu}+\varepsilon^{\nu\alpha}_{\quad\beta}\delta\mathcal G^{\,\mu\beta} \bar\xi_{\nu}+\varepsilon^{\mu\nu\beta}\delta\mathcal {G^{\alpha}}_{\beta} \bar\xi_{\nu}\big\}+\frac{1}{2\mu_{G}}\varepsilon^{\alpha\nu}_{\quad\beta}\delta\mathcal G^{\,\mu\beta} \bar \nabla_{\alpha}\bar\xi_{\nu}\,.
\ee
we could write  last term as $\bar\eta_{\beta}\delta\mathcal G^{\,\mu\beta}$ if we define $\bar\eta_{\beta}\equiv\frac12\,\varepsilon^{\alpha\nu}_{\quad\beta}\,\bar \nabla_{\alpha}\bar\xi_{\nu}$. Again we write the conserved current as a total derivative
\bea \label{current2} 
\mathcal K^{\mu}&=&\frac{1}{\mu_{G}}\bar\nabla_{\nu}\big({\cal F}_1^{\mu\nu}(\bar\eta)+{\cal F}_4^{\mu\nu}\big)\,,\nn\\
{\cal F}_4^{\mu\nu}&=&\frac12 \big\{\varepsilon^{\mu\nu\beta}\delta\mathcal G_{\,\alpha\beta} \bar\xi^{\alpha}+\varepsilon^{\mu\alpha\beta}\delta\mathcal {G^{\nu}}_{\beta} \bar\xi_{\alpha}-\varepsilon^{\nu\alpha\beta}{\delta\mathcal G^{\mu}}_{\beta} \bar\xi_{\alpha}\big\}\,.
\eea
\subsection{Asymptotic behavior of Extremal solutions}
By inserting results in equations (\ref{current1}) and (\ref{current2}) into equation (\ref{cc1}) we can compute conserved charge. Since the conserved current is written as a total derivative, integral over  bulk will be equal to an integral over boundary
\be \label{cc2}
Q(\bar{\xi})=\frac{1}{8\pi G}\int_{\prt{\cal M}} dS_{i}\,\sqrt{-\bar g}\,{\cal F}_{tot}^{0i}=\lim_{\rho\rightarrow\infty} \frac{1}{8\pi G}\int d\phi\,\sqrt{-\bar g}\,{\cal F}_{tot}^{0\rho}\,,
\ee
where
\bea{\cal F}_{tot}^{0\rho}&=&\Big(1+2(D \kappa_1+\kappa_2)\,\Upsilon+3\,(\kappa_3+D\kappa_4+D^2\kappa_5)\,\Upsilon^2\Big)\,{\cal F}_1^{0\rho}(\xi)+\frac{1}{\mu_{G}}{\cal F}_1^{0\rho}(\eta)\nn\\&+&\Big(2\kappa_1+\kappa_2+\big(3\kappa_3+(D+4)\kappa_4+6D\kappa_5\big)\,\Upsilon\Big){\cal F}_2^{0\rho}+\Big(\kappa_2+(3\kappa_3+D\kappa_4)\,\Upsilon\Big){\cal F}_3^{0\rho}+\frac{1}{\mu_{G}}{\cal F}_4^{0\rho}\,.\eea
As mentioned before, extremal solutions (\ref{esol}) are asymptotically $AdS_3$, so we have $SL(2,R)_L\times SL(2,R)_R$ symmetry at boundary  and therefore we have two copies of   Killing vectors, $\xi_{n}^L$ and $\xi_{n}^R$, which generate two copies of  Virasoro algebra
\be 
i[\xi_{m}^L,\xi_{n}^L]=(m-n)\xi_{m+n}^L,\quad i[\xi_{m}^R,\xi_{n}^R]=(m-n)\xi_{m+n}^R,\quad  [\xi_{m}^L,\xi_{n}^R]=0. 
\ee
Going to light-cone coordinates with  $\tau^{\pm}=t\pm\,l\phi$, the background metrics (\ref{esol}) change to
\bea 
ds_{log}^2&=&2(D\rho^{\nu}\,\ln(\rho)+D_0)(d\tau^{-})^2-\frac{2\rho}{l^2}\,d\tau^{+}d\tau^{-}+\frac{l^2 d\rho^2}{4\,\rho^2}\,,\nn\\
ds_{poly}^2&=&2(C\rho^{\nu}+C_0)(d\tau^{-})^2-\frac{2\rho}{l^2}\,d\tau^{+}d\tau^{-}+\frac{l^2 d\rho^2}{4\,\rho^2}\,, \eea
which are asymptotically $AdS_3$ and therefore one can use standard Brown-Henneaux  asymptotically $AdS_3$ boundary conditions \cite{BH}. Since in above metrics  $\rho^{\nu}\,\ln\rho$ and $\rho^{\nu}$ terms are diverging more slowly than $\rho$  at boundary ($0<\nu<1$), we can choose boundary fluctuations as
\be \label{bc}
\left(\begin{array}{ccc}
h_{++}\sim{\cal O}(1) & h_{+-} \sim {\cal O}(1) & h_{+\rho} \sim {\cal O}(\frac{1}{\rho^2}) \\
                         & h_{--} \sim {\cal O}(1)& h_{-\rho} \sim {\cal O}(\frac{1}{\rho^2}) \\
                         &                            & h_{\rho\rho} \sim {\cal O}(\frac{1}{\rho^3})
\end{array}\right).\ee
The most general diffeomorphism which preserves (\ref{bc}) is 
\bea 
\xi &=& \xi^{\mu}\prt_{\mu}=[ \epsilon^{+}(\tau^{+})+\frac{2}{\rho}  \prt_{-}^2 \epsilon^{-}(\tau^{-})+{\cal O}(\frac{1}{\rho^2}) ] \prt_{+}
+[ \epsilon^{-}(\tau^{-})+\frac{2}{\rho}  \prt_{+}^2 \epsilon^{+}(\tau^{+})+{\cal O}(\frac{1}{\rho^2}) ] \prt_{-}\nn\\
&-&\frac12 [ \prt_{+}\epsilon^{+}(\tau^{+})+\prt_{-} \epsilon^{-}(\tau^{-})+{\cal O}(\frac{1}{\rho}) ] \prt_{\rho}\,,
\eea
where left and right moving functions are parametrized  by $\epsilon^{+}(\tau^{+})=e^{im\tau_{+}}$ and $\epsilon^{-}(\tau^{-})=e^{in\tau_{-}}$.
We can parametrize asymptotic boundary conditions (\ref{bc}) in order to have true conserved charges and consistent Lie derivative equations
\bea \label{nbc} 
h_{++}&=&f_{++}(t,\phi)+...\,,\quad h_{+-}=f_{+-}(t,\phi)+...\,,\quad h_{--}=f_{--}(t,\phi)+...\,,\nn\\
h_{+\rho}&=&\frac{1}{\rho^2}f_{+\rho}(t,\phi)+...\,,\quad h_{-\rho}=\frac{1}{\rho^2}f_{+\rho}(t,\phi)+...\,,\quad h_{\rho\rho}=\frac{1}{\rho^3}f_{\rho\rho}(t,\phi)+...\,,
\eea 
where $"..."$ are next leading order terms which do not contribute to conserved charges. Plugging (\ref{nbc}) into (\ref{cc2}) and taking $\rho\rightarrow\infty$ limit, lead to
\bea \label{Qcc} Q&=&\frac{1}{8\pi G l}\,\int d\phi\,\Big\{(1-\frac{1}{2m^2 l^2}-\frac{1}{8m^4 l^4}-\frac{1}{\mu_{G}l})\epsilon^{+} f_{++}+(1-\frac{1}{2m^2 l^2}-\frac{1}{8m^4 l^4}+\frac{1}{\mu_{G}l})\epsilon^{-} f_{--}\nn\\
&&\quad\quad\quad-\,(1-\frac{1}{2m^2 l^2}-\frac{1}{8m^4 l^4})\,\big(f_{+-}-\frac{2}{l^4}f_{\rho\rho}\big)(\epsilon^{+}+\epsilon^{-})+\frac{3}{2\mu l}\big(f_{+-}-\frac{2}{l^4}f_{\rho\rho}\big)(\epsilon^{+}-\epsilon^{-})\Big\}\,,\eea
where we have inserted values of $\kappa_1,...,\kappa_5$.
If we insert boundary conditions (\ref{nbc}) into  linearized equations of motion (\ref{LEOM}) then the $\rho\rho$ component at  $\rho\rightarrow\infty$, will give an asymptotic constraint for equation (\ref{Qcc})
\be f_{+-}-\frac{2}{l^4}f_{\rho\rho}=0\,.\ee
According to this constraint the second line of (\ref{Qcc}) becomes zero at boundary. Now we can define left and right moving conserved charges as
\bea 
&&Q_L=\frac{1}{8\pi G l}\,\int d\phi\,(1-\frac{1}{2m^2 l^2}-\frac{1}{8m^4 l^4}-\frac{1}{\mu_{G}l})\,\epsilon^{+}f_{++}\,,\nn\\
&&Q_R=\frac{1}{8\pi G l}\,\int d\phi\,(1-\frac{1}{2m^2 l^2}-\frac{1}{8m^4 l^4}+\frac{1}{\mu_{G}l})\,\epsilon^{-}f_{--}\,,
\eea
where $Q=Q_L+Q_R$. These charges satisfy two copies of Virasoro algebra with left and right moving central charges
\be \label{fcc1} c_L=\frac{3l}{2G}(1-\frac{1}{2m^2 l^2}-\frac{1}{8m^4 l^4}-\frac{1}{\mu_{G}l})\,,\quad c_R=\frac{3l}{2G}(1-\frac{1}{2m^2 l^2}-\frac{1}{8m^4 l^4}+\frac{1}{\mu_{G}l})\,.
\ee
One can easily check that  $\mu_{G}\rightarrow\infty$ is consistent with central function formalism in \cite{Kraus:2005vz}, i.e.
\be 
c=\frac{l}{2G}\,g_{\mu\nu}\,\frac{\prt{\cal L}}{\prt R_{\mu\nu}}\,,
\ee
so simply we can read central charges of the dual CFT to the asymptotically $AdS$ sector of Born-Infeld Lagrangian
\be 
\label{fcc2} c_L=\frac{3l}{2G}(\sqrt{1-\frac{1}{m^2 l^2}}-\frac{1}{\mu_{G}l})\,,\quad c_R=\frac{3l}{2G}(\sqrt{1-\frac{1}{m^2 l^2}}+\frac{1}{\mu_{G}l})\,.
\ee
As we see, the left central charges we have found here in (\ref{fcc1}) and (\ref{fcc2}), are exactly those in table 11, which we found from  Cardy's formula.

There is an important point to note. As we see, although we have ignored all terms which contain field strength of the gauge field, final results for central charges have not changed. We can check that these extra terms fall off more rapidly than pure gravity terms as one goes to boundary $(\rho\rightarrow\infty)$. As an example consider the $F^2$ term in Lagrangian. The contribution of this term to equations of motion is as follow
\be
T^{(em)}_{\mu\nu}=-\frac12 g_{\mu\nu} F^2+2 {F_\mu}^\alpha F_{\alpha\nu}\,,
\ee
which leads to the following electromagnetic conserved current
\be
\mathcal K^{\mu}_{(em)}=-\frac12\xi_\nu h^{\mu\nu} F^2+\xi^\mu h^{\alpha\beta} {F_\alpha}^\rho F_{\rho\beta}-2\xi_\nu h^{\alpha\beta} {F^\mu}_\alpha {F_\beta}^\nu\,.
\ee
For both logarithmic and polynomial solutions the first term above is zero because of $F^2=0$. One can see that as we go to boundary $(\rho\rightarrow\infty)$ the next two terms also vanish. This can be checked by choosing boundary fluctuations (\ref{bc}) and using the fact that our solutions are restricted by $(0<\nu<1)$. The same behavior still holds for all other terms which contain gauge field strength.\footnote{For a similar argument in four dimensions see for example \cite{Garousi:2009zx}}
\section{Asymptotic conserved charges for non-extremal solutions}
For asymptotically $AdS_3$ metric we used useful relations in equation (\ref{adsback}) for linearization of  equations of motion. But for asymptotically warped-$AdS_3$ metric we do not have these properties and we must linearize  equations of motion around an arbitrary general background metric.
As an example let us start from  second order derivative terms in Lagrangian. Most of calculations are similar to those performed in \cite{Compere:2008cv}. For an action including Einstein-Hilbert and TMG terms we have 
\be 
R_{\mu\nu}-\frac12 g_{\mu\nu}\,R+\Lambda\,g_{\mu\nu}+\frac{1}{\mu_G}\,C_{\mu\nu}=0\,.
\ee
If we define ${\cal G}^{\mu\nu}=R^{\mu\nu}-\frac12 g^{\mu\nu}\,R+\Lambda\,g^{\mu\nu}$ then the conserved current associated to Einstein-Hilbert terms will be
\bea 
\xi_{\nu}\,\delta {\cal G}^{\mu\nu}&=&\xi_{\nu}\Big(-2\,\bar R^{\mu(\alpha}{h_{\alpha}}^{\nu)}+\frac12\,(2\bar\nabla_{\alpha}\bar\nabla^{(\mu}h^{\nu)\alpha}-\bar\Box h^{\mu\nu}-\bar\nabla^{\mu}\bar\nabla^{\nu}h)+\frac12(\bar R-2\Lambda)h^{\mu\nu}\Big)\nn\\
&-&\frac12 \xi^{\mu}\Big(-h^{\alpha\beta}\bar R_{\alpha\beta}+\bar\nabla_{\alpha}\bar\nabla_{\beta}h^{\alpha\beta}-\bar\Box h\Big)\,.
\eea
Using the Killing vector equation and $\bar\nabla^{\alpha}\bar\nabla_{\beta}\bar \xi_{\nu}={\bar R^{\mu}}_{\nu\alpha\beta} \bar \xi_{\mu}$ this current can be expressed as \cite{Bouchareb:2007yx}
\bea 
\label{KEH} 
{\cal K}_{EH}^{\mu}&=&\bar\nabla_{\alpha}{\cal F}_{EH}^{\mu\alpha}-\xi^{\nu}{\cal G}^{\mu\alpha} h_{\alpha\nu}+\frac12\,\xi^{\mu}{\cal G}^{\alpha\nu} h_{\alpha\nu}-\frac12\,\xi^{\nu}{{\cal G}^{\mu}}_{\nu} h\,,\\
{\cal F}_{EH}^{\mu\nu}&=&\frac12\,\Big\{ \xi^{\nu}\nabla_{\alpha}h^{\mu\alpha}-\xi^{\mu}\nabla_{\alpha}h^{\nu\alpha}+\xi_{\alpha}\nabla^{\mu} h^{\alpha\nu}-\xi_{\alpha}\nabla^{\nu} h^{\alpha\mu}\nn\\
&+&\xi^{\mu}\nabla^{\nu}h-\xi^{\nu}\nabla^{\mu}h+h^{\alpha\nu}\nabla_{\alpha}\xi^{\mu}-h^{\alpha\mu}\nabla_{\alpha}\xi^{\nu}+h \nabla^{[\mu}\xi^{\nu]}\Big\}\,.
\eea
For the TMG part the associated conserved charge is given by 
\be
\xi_{\nu}\delta C^{\mu\nu}=\frac{1}{2\sqrt{g}}\,\xi_{\nu}\Big\{\epsilon^{\mu\alpha\beta}\,\nabla_{\alpha}\delta {G^{\nu}}_{\beta}+\epsilon^{\nu\alpha\beta}\,\nabla_{\alpha}\delta {G^{\mu}}_{\beta}+\epsilon^{\mu\alpha\beta}\,\delta\Gamma^{\nu}_{\alpha\lambda} {G^{\lambda}}_{\beta}+\epsilon^{\nu\alpha\beta}\,\delta\Gamma^{\mu}_{\alpha\lambda} {G^{\lambda}}_{\beta}-\sqrt{g} h C^{\mu\nu}\Big\}\,.
\ee
Again if we define $\eta^{\nu}\equiv\frac{1}{2\sqrt{g}}\epsilon^{\nu\alpha\beta}\nabla_{\alpha}\xi_{\beta}$  we can find the following equations \cite{Bouchareb:2007yx}
\bea 
{\cal K}_{CS}^{\mu}&=&\bar\nabla_{\alpha}{\cal F}_{CS}^{\mu\alpha}(\xi)-\xi^{\nu}C^{\mu\alpha} h_{\alpha\nu}+\frac12\,\xi^{\mu}C^{\alpha\nu} h_{\alpha\nu}-\frac12\,\xi^{\nu}{C^{\mu}}_{\nu} h\,,\\ \label{KCS} 
{\cal F}_{CS}^{\mu\nu}(\xi)&=&{\cal F}_{EH}^{\mu\nu}(\eta)+\frac{1}{\sqrt{g}}\,\xi_{\lambda}\big(\epsilon^{\mu\nu\alpha}\,\delta {G^{\lambda}}_{\alpha}-\frac12\,\epsilon^{\mu\nu\lambda} \delta G\big)\nn\\ 
&+&\frac{1}{2\sqrt{g}}\epsilon^{\mu\nu\alpha}\big[\xi_{\alpha} h^{\lambda\beta} G_{\lambda\beta}+\frac12\,h\,(\xi^{\beta} G_{\beta\alpha}+\frac12 \xi_{\alpha} R)\big]\,.
\eea
The sum of the last three terms of equation (\ref{KEH}) and equation (\ref{KCS}) is zero by using equations of motion. Therefore conserved charge becomes
\be \label{WCCQ} Q=\frac{1}{8\pi G}\int_{\cal M} \sqrt{g}\,{\cal K}^{\mu}=\frac{1}{8\pi G}\int_{\prt\cal M}\,\sqrt{g}\,\Big({\cal F}_{EH}^{0i}(\xi)+\frac{1}{\mu_{G}}{\cal F}_{CS}^{0i}(\xi)\Big)\,d S_{i}\,.\ee
For warped solution the metric is 
\be 
ds^2=-\beta^2\,\frac{\rho^2-\rho_0^2}{r^2}\,dt^2+r^2\big(d\phi-\frac{\rho+(1-\beta^2)\,\omega}{r^2}\,dt\big)^2+\frac{1}{\beta^2 \zeta^2}\,\frac{d\rho^2}{\rho^2-\rho_0^2}\,,
\ee
and  asymptotic boundary fluctuations can be defined as
\be \label{Wbc}
\left(\begin{array}{ccc}
h_{tt}\sim{\cal O}(\frac{1}{\rho}) & h_{t\phi} \sim {\cal O}(1) & h_{t\rho} \sim {\cal O}(\frac{1}{\rho^2}) \\
                         & h_{\phi\phi} \sim {\cal O}(\rho)& h_{\phi\rho} \sim {\cal O}(\frac{1}{\rho}) \\
                         &                            & h_{\rho\rho} \sim {\cal O}(\frac{1}{\rho^3})
\end{array}\right)\,.\ee
The most general diffeomorphism which preserves (\ref{Wbc}) is
\be \label{diffwarp}
\xi={\cal N} \epsilon(\phi)\,\prt_{t}-\rho\,\epsilon'(\phi)\,\prt_{\rho}+\epsilon(\phi)\,\prt_{\phi}\,,
\ee
where prime is derivative with respect to $\phi$  and ${\cal N}$ is an arbitrary constant which does not affect our results\cite{Compere:2007in}. All remaining steps for finding central charge is similar to the previous case. Since $\phi$ is a periodic coordinate with $\phi\sim\phi+2\pi$ then it is better to use the Fourier analysis by considering  $\epsilon(\phi)=e^{i n \phi}$ in (\ref{diffwarp}). Inserting this vector into equation (\ref{WCCQ}) we will find exactly the left central charge in table 10 i.e. 
\bea 
c_L=\frac{12\pi}{\kappa \beta^2}(2+\frac{2\beta^2-1}{\mu_{G}})=3.2^6\pi\frac{z}{(1-2z)^2}\Xi_1\,.
\eea
As before we can show that at this order of calculations conserved current associated to gauge field Lagrangian $F^2$, falls-off more rapidly than  gravitational terms at boundary and it has not any contribution to central charge.

Although this approach confirms left central charges found by Cardy's formula, it is unable to find  right central charge, see \cite{Compere:2008cv} and \cite{Compere:2007in} for the same obstruction. This is because the asymptotic symmetry is $SL(2,R)\times U(1)$  and we must expect that since we have one $SL(2,R)$, we will find one of the central charges. It will be interesting to find a way to compute the other central charges of this algebra. This situation still holds in presence of higher order curvature terms with more lengthy calculations and again we can only confirm  left central charges of table 10. 

\section{Summary and Conclusions}
In this paper we have investigated several solutions corresponding to  three dimensional extended massive theories of gravity. The extension has been done either by adding  gauge fields through a Maxwell term or Maxwell-Chern-Simons term or by adding a gravitational Chern-Simons term. In all these cases we have found charged solutions.

These charged solutions are as follows:

1. Extremal logarithmic and polynomial self dual black holes with and without gravitational Chern-Simons term.

2. Maxwell charged solutions which have naked singularities.

3. Non-extremal warped-$AdS$ solutions.

We have found these solutions for different orders of expansion (up to six) of Born-Infeld Lagrangian  and for unexpanded Born-Infeld Lagrangian. The results of logarithmic solutions are given in  tables 3,4,7 and 8 while results for polynomial solutions are given in the tables 2 and 6. 
Values of cosmological constant for extremal solutions are the same at each level of expansion and are given in table 1. 

Results for Maxwell-charged solutions  are given in table 5.
Results for non-extremal warped-$AdS$ solutions are given in tables 9 and 10. Values of cosmological constant differ from those of extremal cases.

For all solutions we have found  super-angular momentum as a conserved current for $SL(2,R)$ symmetry. Then we have read  mass and angular momentum. For extremal cases we always find  $J=Ml$ or $\Delta=0$ but for  non-extremal case we find (\ref{nonjm}) or $\Delta=2\pi(\delta{\bs J}^Y)$.

Results for mass have been checked by consistency between the first law of thermodynamics for black holes and the Smarr-like formula. For extremal solutions Smarr-like formula is $M=\frac12\,T_H S\,+\Omega_h J+\,\Phi_h \bar{Q}$ but for non-extremal solution it is given by $M=T_H S\,+2\Omega_h J+\frac12\,\Phi_h \bar{Q}$.

We have also found entropy of each black hole, by using  (\ref{ent1}) when we have no gravitational Chern-Simons term and we have used (\ref{ent2}) when we have it. The value of entropy helps us to read central charges of  dual conformal field theories. 

In this paper we used Cardy's formula to find  central charges.
For asymptotically $AdS$ solutions we can only find  left central charge (table 11) of dual CFTs from  Cardy's formula. 
For asymptotically warped-$AdS$ solutions the dual CFT  is different from asymptotically $AdS$ solutions. 
The difference between left and right central charge gives the value of holographic gravitational anomaly. Our results show that the value of anomaly only depends on the coupling of gravitational Chern-Simons term and is independent of expansion of  Born-Infeld Lagrangian.

To confirm our results for central charges of dual CFTs we have used another approach. We calculate conserved charges associated to asymptotic symmetry transformations of solutions, i.e. $SL(2,R)\times SL(2,R)$ for asymptotically $AdS$ solutions and $SL(2,R)\times U(1)$ for asymptotically warped-$AdS$ ones.
By choosing a proper gravitational perturbation at boundaries we can find exactly central charges that we have found by  Cardy's formula. By this approach we can compute both left and  right central charges of the CFT dual to asymptotically $AdS$ solutions. But for dual CFTs associated to asymptotically warped-$AdS$ solutions  only  left central charges can be computed.

\appendix
\section{Useful relations for linearization}
In order to study  fluctuations of a generic action around some background  we need to expand various tensors up to second order in metric perturbations $h_{\mu\nu}$. Using 
\be 
g_{\mu\nu}\equiv\bar g_{\mu\nu}+h_{\mu\nu},
\ee
linearized Christoffel symbols and  Riemann, Ricci and scalar tensors become 
\bea 
\delta\Gamma^{\mu}_{\nu\rho}&=&\frac12\,\bar g^{\mu\lambda}(\bar\nabla_{\nu}h_{\lambda\rho}+\bar\nabla_{\rho}h_{\lambda\nu}-\bar\nabla_{\lambda}h_{\nu\rho})\,,\nn\\
\delta R^{\mu}_{\nu\rho\sigma}&=&\frac12(\bar\nabla_{\rho}\bar\nabla_{\sigma}h^{\mu}_{\,\,\nu}+\bar\nabla_{\rho}\bar\nabla_{\nu}h^{\mu}_{\,\,\sigma}-\bar\nabla_{\rho}\bar\nabla^{\mu}h_{\sigma\nu}-\bar\nabla_{\sigma}\bar\nabla_{\rho}h^{\mu}_{\,\,\nu}-\bar\nabla_{\sigma}\bar\nabla_{\nu}h^{\mu}_{\,\,\rho}+\bar\nabla_{\sigma}\bar\nabla^{\mu}h_{\rho\nu})\,,\nn\\
\delta R_{\mu\nu}&=&\frac12\,(\bar\nabla^{\lambda}\bar\nabla_{\mu}h_{\lambda\nu}+\bar\nabla^{\lambda}\bar\nabla_{\nu}h_{\lambda\mu}-\bar\Box h_{\mu\nu}-\bar\nabla_{\mu}\bar\nabla_{\nu}h)\,,\nn\\
\delta R&=&(\bar\nabla^{\lambda}\bar\nabla^{\rho}h_{\lambda\rho}-\bar\Box h\,)-\frac{2\Lambda_0}{D-2}\,h\,.
\eea
These relations help us to linearize different terms in  equations of motion. By defining $\Upsilon=\frac{2\Lambda_0}{D-2}$ we find the following relations
\bea \label{appA1}
&&\delta(R_{\mu\rho\nu\sigma}R^{\rho\sigma})=\frac{1}{D-1}\big((D-2)\Upsilon \delta R_{\mu\nu}+\Upsilon \bar g_{\mu\nu} \delta R+\Upsilon^2 h_{\mu\nu}\big) \,,\quad\delta(RR_{\mu\nu})=D\Upsilon \delta R_{\mu\nu}+\Upsilon\bar g_{\mu\nu} \delta R \,,\nn\\
&&\delta(R_{\mu}^{\alpha}R_{\alpha\nu})=2\Upsilon \delta R_{\mu\nu}-\Upsilon^2h_{\mu\nu} \,,\quad \delta(R_{\mu\nu}R^{\mu\nu})=2\Upsilon \delta R \,,\quad
\delta(R_{\mu}^{  \alpha}R_{\alpha\beta}R^{\beta}_{  \nu})=3\Upsilon^2 \delta R_{\mu\nu}-2\Upsilon^3 h_{\mu\nu} \,,\nn\\
&&\delta(g_{\mu\nu} R_{\alpha\beta}R^{\beta\rho}R_{\rho}^{  \alpha})=3\Upsilon^2 \bar g_{\mu\nu} \delta R^L+D\Upsilon^3 h_{\mu\nu}  \,,\quad
\delta(R_{\mu\nu} R_{\alpha\beta} R^{\alpha\beta})=D\Upsilon^2 \delta R_{\mu\nu}+2\Upsilon^2 \bar g_{\mu\nu} \delta R  \,,\nn\\
&&\delta(R R_{\mu}^{\alpha}R_{\alpha\nu})=2D\Upsilon^2 \delta R_{\mu\nu}+\Upsilon^2 \bar g_{\mu\nu} \delta R-D\Upsilon^3 h_{\mu\nu} \,,\quad
\delta(g_{\mu\nu}R R_{\alpha\beta} R^{\alpha\beta} )=3D\Upsilon^2 \bar g_{\mu\nu} \delta R+D^2 \Upsilon^3 h_{\mu\nu} \,,\nn\\
&& \delta(g_{\mu\nu} R^3)=3D^2 \Upsilon^2 \bar g_{\mu\nu} \delta R-D^3\Upsilon^3 h_{\mu\nu}\,,\\
&& \delta(\Box R_{\mu\nu})=\bar\Box \delta R_{\mu\nu}-\Upsilon \bar\Box h_{\mu\nu} \,,\quad \delta(\nabla_{\mu}\nabla_{\nu} R_{\alpha\beta})=\bar\nabla_{\mu}\bar\nabla_{\nu} \delta R_{\alpha\beta}-\Upsilon \bar\nabla_{\mu}\bar\nabla_{\nu} h_{\alpha\beta} \,,\quad
\delta(\Box R)=\bar\Box \delta R \,,\nn\\
&&\delta(\nabla_{\mu}\nabla_{\nu} R)=\bar\nabla_{\mu}\bar\nabla_{\nu} \delta R \,,\quad \delta(\nabla_{\alpha} \nabla_{\beta}R^{\alpha\rho}R^{   \beta}_{\rho})=2\Upsilon \bar\nabla_{\alpha} \bar\nabla_{\beta}\delta R^{\alpha\beta}-2\Upsilon^2 \bar\nabla_{\alpha} \bar\nabla_{\beta}h^{\alpha\beta} \,,\nn\\
&& \delta(\Box R_{\mu}^{\alpha}R_{\alpha\nu})=2\Upsilon \bar\Box \delta R_{\mu\nu}-2\Upsilon^2 \bar\Box h_{\mu\nu}\,,\quad 
\delta(\nabla_{\alpha}\nabla_{(\mu}R_{\nu)\beta}R^{\beta\alpha})=2\Upsilon\bar\nabla^{\alpha}\bar\nabla_{(\mu} \delta R_{\nu)\alpha}-2\Upsilon^2 \bar\nabla^{\alpha}\bar\nabla_{(\mu} h_{\nu)\alpha}\,,\nn\\
&& \delta(\nabla_{\alpha} \nabla_{\beta}R^{\alpha\beta}R)=D\Upsilon(\bar\nabla_{\alpha} \bar\nabla_{\beta}\delta R^{\alpha\beta}-\Upsilon \bar\nabla_{\alpha} \bar\nabla_{\beta}h^{\alpha\beta})+\Upsilon \bar\Box \delta R\,,\quad  \delta(\nabla_{\mu}\nabla_{\nu} [R_{\alpha\beta}R^{\alpha\beta}])=2\Upsilon \bar\nabla_{\mu}\bar\nabla_{\nu}\delta R\,,\nn\\
&&\delta(\nabla_{\alpha}\nabla_{(\mu}R_{\nu)}^{\alpha}R)=\Upsilon (\bar\nabla_{\alpha}\bar\nabla_{(\mu}\delta R_{\nu)}^{\alpha}-\Upsilon \bar\nabla_{\alpha}\bar\nabla_{(\mu} h_{\nu)}^{\alpha})+\Upsilon \bar\nabla_{(\mu}\bar\nabla_{\nu)}\delta R\,,\quad \delta(\Box [R_{\alpha\beta}R^{\alpha\beta}])=2\Upsilon \bar\Box \delta R\,,\nn\\
&&\delta(\Box R R_{\mu\nu})=D\Upsilon(\bar\Box \delta R_{\mu\nu}-\Upsilon \bar\Box h_{\mu\nu})+\Upsilon \bar g_{\mu\nu} \bar\Box \delta R\,,\quad
\delta([g_{\mu\nu} \Box-\nabla_{\mu} \nabla_{\nu}] R^2)=2D \Upsilon (\bar g_{\mu\nu} \bar\Box-\bar\nabla_{\mu} \bar\nabla_{\nu}) \delta R\,.\nn \\
\label{appA2}
\eea
\section*{Acknowledgment}
This work was supported by Ferdowsi University of Mashhad under the grant 3/18769 (30/08/1390). A. G. would like to thank Amir Esmaeil Mosaffa for useful comments and reading the manuscript.

 \end{document}